\documentclass{article}

\usepackage{arxiv}

\usepackage[utf8]{inputenc} 
\usepackage[T1]{fontenc}    
\usepackage{hyperref}       
\usepackage{url}            
\usepackage{booktabs}       
\usepackage{amsfonts}       
\usepackage{nicefrac}       
\usepackage{microtype}      
\usepackage{lipsum}		
\usepackage{graphicx}
\usepackage{natbib}
\usepackage{doi}

\usepackage{xspace}
\usepackage{pifont}       
\usepackage{booktabs}     
\usepackage{graphicx}     
\usepackage{array}        
\usepackage{tabularx}     
\usepackage{amsmath}      
\usepackage{amssymb}      
\usepackage{amsthm}       
\usepackage{snotez}      
\usepackage{mathtools}
\usepackage{adjustbox}
\usepackage{multirow}
\usepackage{url}          
\usepackage{listings}
\usepackage{float}                
\newfloat{featurebox}{htbp}{lob}  
\floatname{featurebox}{Box}
\usepackage[most]{tcolorbox}
\tcbuselibrary{listings,breakable}

\usepackage{algorithm}
\usepackage{algpseudocode}

\usepackage{multibib}
\newcites{supp}{Supplementary References} 

\definecolor{customgreen}{RGB}{0, 150, 0}

\newcommand{\cmark}{\textcolor{customgreen}{\ding{51}}} 
\newcommand{\xmark}{\textcolor{red}{\ding{55}}}

\def\ours{\text{CellMaster}\xspace}

\newcommand{\xhdr}[1]{\noindent{{\bf #1.}}}

\newcommand{\hide}[1]{}

\graphicspath{{figures/}} 

\theoremstyle{thmstyleone}

\theoremstyle{thmstyletwo}

\hypersetup{
    colorlinks=true,
    linkcolor=black,
    filecolor=black,      
    urlcolor=black,
    citecolor=black
}

\graphicspath{{figures/}}
\def\ours{\text{CellMaster}\xspace}

\title{CellMaster: 
Collaborative Cell Type Annotation in Single-Cell Analysis}

\author{
  Zhen Wang\textsuperscript{1,$\dagger$,$\ast$}, 
  Yiming Gao\textsuperscript{2,$\dagger$}, 
  Jieyuan Liu\textsuperscript{1,$\dagger$}, 
  Enze Ma\textsuperscript{1}, 
  Jefferson Chen\textsuperscript{1}, 
  Mark Antkowiak\textsuperscript{5}, \\
  \textbf{Mengzhou Hu}\textsuperscript{3}, 
  \textbf{JungHo Kong}\textsuperscript{3}, 
  \textbf{Dexter Pratt}\textsuperscript{3}, 
  \textbf{Zhiting Hu}\textsuperscript{1}, 
  \textbf{Wei Wang}\textsuperscript{4}, 
  \textbf{Trey Ideker}\textsuperscript{3}, 
  \textbf{Eric P. Xing}\textsuperscript{6,7} \\
  \vspace{2mm} \\
  \fontsize{9}{11}\selectfont
  \textsuperscript{1}Halicioglu Data Science Institute, University of California, San Diego, CA, USA \\
  \textsuperscript{2}Department of Electrical \& Computer Engineering, Texas A\&M University, College Station, TX, USA \\
  \textsuperscript{3}Department of Medicine, University of California, San Diego, CA, USA \\
  \textsuperscript{4}Department of Chemistry and Biochemistry, University of California San Diego, La Jolla, CA, USA \\
  \textsuperscript{5}Moores Cancer Center, University of California, San Diego, La Jolla, CA, USA \\
  \textsuperscript{6}Mohamed bin Zayed University of AI, Abu Dhabi, UAE \\
  \textsuperscript{7}School of Computer Science, Carnegie Mellon University, Pittsburgh, PA, USA
}

\renewcommand{\shorttitle}{\textit{arXiv} Template}

\hypersetup{
pdftitle={A template for the arxiv style},
pdfsubject={q-bio.NC, q-bio.QM},
pdfauthor={David S.~Hippocampus, Elias D.~Striatum},
pdfkeywords={First keyword, Second keyword, More},
}

\begin{document}
\maketitle

{
  \renewcommand{\thefootnote}{} 
  \footnotetext{$^\ast$Corresponding author: \texttt{zhw085@ucsd.edu}}
  \footnotetext{$^\dagger$These authors contributed equally to this work.}
}

\begin{abstract}
Single-cell RNA-seq (scRNA-seq) enables atlas-scale profiling of complex tissues, revealing rare lineages and transient states. Yet, assigning biologically valid cell identities remains a bottleneck because markers are tissue- and state-dependent, and novel states lack references.
We present CellMaster, an AI agent that mimics expert practice for zero-shot cell-type annotation. Unlike existing automated tools, CellMaster leverages LLM-encoded knowledge (e.g., GPT-4o) to perform on-the-fly annotation with interpretable rationales, without pre-training or fixed marker databases. Across 9 datasets spanning 8 tissues, CellMaster improved accuracy by 7.1\% over best-performing baselines (including CellTypist and scTab) in automatic mode. With human-in-the-loop refinement, this advantage increased to 18.6\%, with a 22.1\% gain on subtype populations. The system demonstrates particular strength in rare and novel cell states where baselines often fail.
Source code and the web application are available at \href{https://github.com/AnonymousGym/CellMaster}{https://github.com/AnonymousGym/CellMaster}.
\end{abstract}


\section{Introduction}

Single-cell RNA-seq (scRNA-seq) has transformed the study of complex tissues by resolving closely related lineages and transient cellular states within their native contexts. Tissues such as PBMC and developmental liver contain fine-grained subtypes (e.g., Classical vs Non-classical monocytes; Regulatory vs Memory T cells) and short-lived intermediates (hepatoblasts; intermediate mature Neutrophils; pre-B cells) whose marker usage is highly context-dependent. For example, while Alb, Cd79a, and S100a8 serve as canonical markers for hepatocytes, B cells, and neutrophils, respectively, their effectiveness is dataset-dependent and not uniformly transferable. As single-cell atlases now span tens of millions of cells and over one hundred tissues ~\cite{Shi2025WholeBodyMap, Zhang2024HUSCH}, scalable and biologically faithful cell type annotation has become a central bottleneck for downstream analysis, interpretation, and cross-study integration.

The classic approach relies on unsupervised clustering by human experts, finding differentially expressed marker genes for each cluster, and assigning identities by analyzing canonical and tissue-specific markers within developmental or perturbation context ~\cite{luecken2019current, tabula2020single, dominguez2022cross}. While this produces high-quality labels, it is time-consuming, scales poorly to datasets, and becomes especially challenging for rare populations and transitional or novel states that lack consensus markers. The definition of cellular identities ultimately relies on expert interpretation to connect molecular signatures with true biological insights.

Automated approaches reduce this burden through marker-table matching (e.g., CellMarker 2.0~\cite{cellmarker2.0}), reference label transfer (e.g., Seurat), or supervised classification (e.g., CellTypist~\cite{celltypist}, scTab~\cite{Fischer2024scTab}). Despite strong performance on common types, these methods rely on reference coverage and nomenclature alignment, which are sensitive to batch-, perturbation-, and domain-shifts, and tend to abstain or mis-assign when confronted with rare or truly novel states, e.g., CellTypist shows especially low performance when annotating clusters with \textless 100 cells ~\cite{dominguez2022cross}. Crucially, their decisions are often difficult to audit biologically, limiting trust in exploratory settings.

Recently, large language models (LLMs) have shown promise in performing cell annotation tasks. It has been found that GPT-4 can accurately assign cell type labels given a list of differential genes for each cell type~\cite{gptcelltype}. Moreover, human feedback on LLM-based predictions can improve cell type prediction, especially for certain biological contexts (e.g. tissue type) where human expertise is required~\cite{agent_in_the_loop_2025, wu2025scextract}. Such human-in-the-loop architectures have been explored for annotation, labeling, and expert-guided model refinement tasks in both general AI and biological domains. For instance, agentic frameworks that combine LLM reasoning with human oversight have been proposed to distill expert knowledge into artificial intelligence models \cite{agent_in_the_loop_2025}, while domain-specific implementations such as scExtract \cite{wu2025scextract} integrate LLM-based retrieval and annotation workflows with expert-curated validation to enhance single-cell data interpretation.

However, a task-grounded system for scRNA-seq that (a) reasons over markers in situ, (b) explains its decisions in biologically interpretable terms, and (c) operates zero-shot, i.e., without pretraining on task-specific labels or reliance on fixed marker tables, has not been established for cell-type annotation.

\begin{figure}[p] 
    \centering    
    \includegraphics[width=\linewidth]{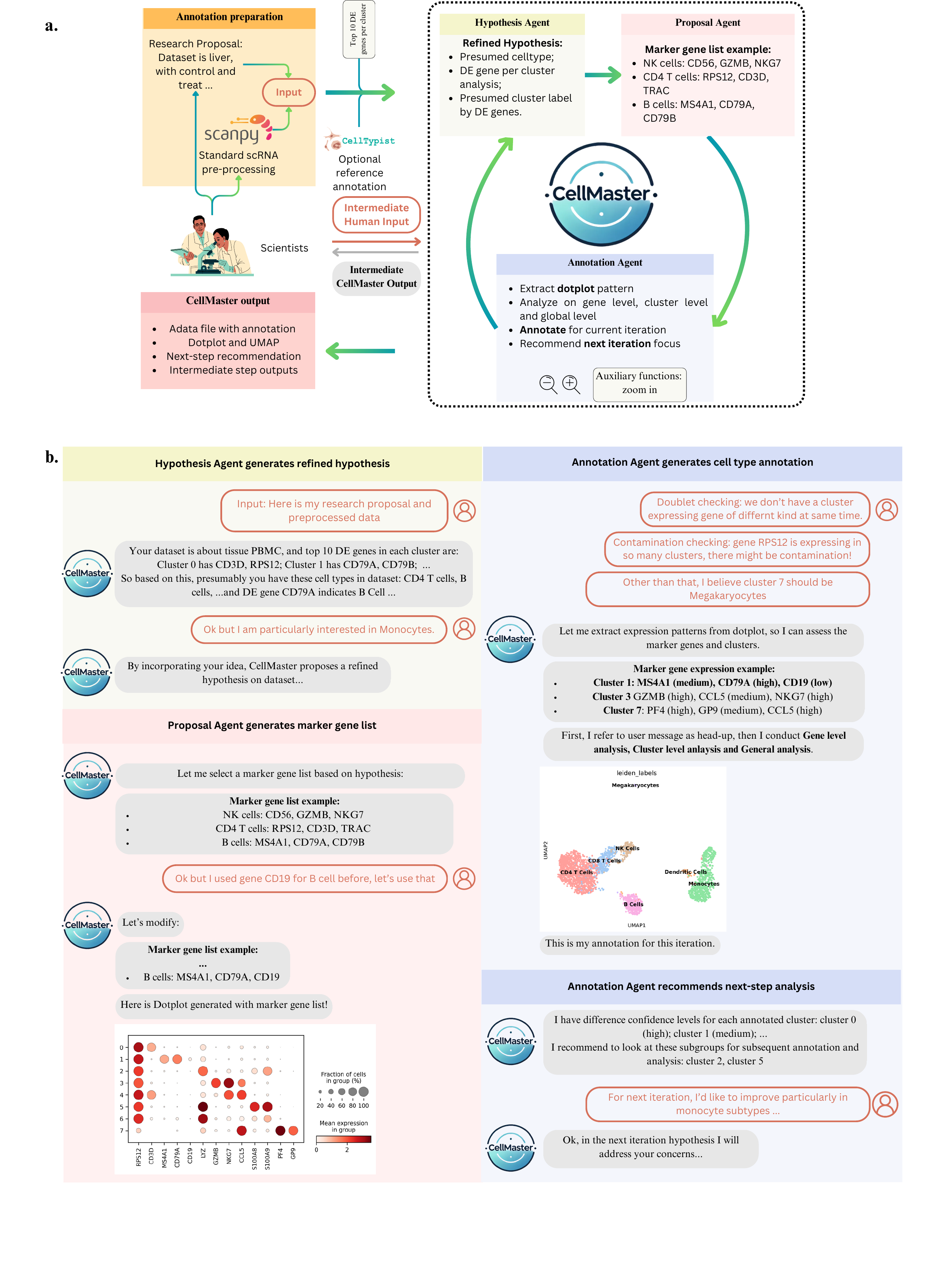}
    \vspace{-50pt}    \caption{\textbf{Overview of \ours architecture and workflow.} 
    \textbf{(a)} System pipeline: three specialized agents (Hypothesis, Proposal, Annotation) coordinate to iteratively refine cell type annotations from user-provided data and context. 
    \textbf{(b)} Example human-AI dialogue showing iterative annotation refinement on a PBMC dataset.}
    \label{fig:workflow}
\end{figure}

To address this gap, we introduce \ours, an LLM-driven agent that mirrors an expert's reasoning loop to deliver zero-shot, explainable cell-type labels by interpreting cluster-level DE markers together with dataset context and producing natural-language rationales without pretraining or fixed marker tables. The system performs biology-aware refinement, iteratively proposing and testing hypotheses to resolve near-neighbor confusions (e.g., NK vs T cells) and escalating uncertain clusters for review, while maintaining sensitivity to rare and candidate-novel states. In addition, a collaborative user interface (UI) allows experts to co-pilot the process, reviewing rationales, suggesting corrections, adjusting granularity, and leaving provenance-tracked comments, so that human judgment directly shapes final labels. We evaluated \ours across 9 datasets spanning 8 tissues, comparing against CellTypist, scTab, CellMarker 2.0, GPTCelltype and Biomni using average CL score with expert adjudication. \ours operationalizes the design principle of emulating human expert workflows with transparent reasoning and structured uncertainty handling to accelerate biologically faithful annotation in evolving atlases (Figure~\ref{fig:workflow}).

\section{Related Works}
\label{sec:why}

The application of LLMs in biomedicine has progressed rapidly, from domain-specific text encoders such as BioGPT~\cite{biogpt} to general-purpose reasoning models like GPT-4o. In single-cell analysis, early efforts followed two largely separate directions. The first comprises foundation models, including scBERT~\cite{scBert}, scGPT~\cite{cui2023scGPT}, and Geneformer~\cite{GeneFormer}, which model genes as tokens to learn latent representations for tasks such as imputation and perturbation prediction. The second focuses on language-based interfaces, such as Cell2Sentence~\cite{rizvi2025scaling} and scChat~\cite{scchat}, which enable natural-language querying of datasets and metadata.

Despite their promise, both paradigms exhibit key limitations. Embedding-based models operate in opaque latent spaces that hinder interpretability, while language interfaces often function as thin wrappers around predefined execution pipelines, offering limited reasoning or adaptability. To bridge this gap, recent work has shifted toward \emph{autonomous agents} that combine LLM reasoning with tool use. General biomedical agents like Biomni~\cite{huang2025biomni} demonstrate broad problem-solving capabilities, while single-cell–specific systems such as scAgent~\cite{mao2025scagentuniversalsinglecellannotation} and scPilot~\cite{gao2025scpilot} formalize end-to-end analytic workflows.

Within this landscape, CASSIA~\cite{cassia} introduces a multi-agent assembly line (Annotator, Validator, Scorer) for automated cell type annotation, and scRAG~\cite{scRAG} improves cross-tissue generalization by augmenting LLM reasoning with retrieved structured knowledge (e.g., cell–tissue relationships). However, these systems primarily emphasize automated accuracy and typically operate as closed pipelines, delivering finalized results with limited opportunity for user intervention. Similarly, scPilot emphasizes systematic automation, while Biomni functions as a comprehensive but heavyweight solver.

In contrast, \textbf{CellMaster} is designed as a \textbf{lightweight, collaborative assistant} that prioritizes iterative human-in-the-loop interaction. Rather than producing a fixed output, CellMaster allows users to intervene during analysis, adjusting annotation granularity (e.g., subdividing broad cell types), critiquing intermediate reasoning, or refining visual interpretations. This open-loop design transforms the LLM from a black-box automation engine into a responsive research partner, aligning computational reasoning with expert biological intuition and hypothesis-driven exploration—capabilities that are increasingly essential for evaluating agents on the rediscovery of scientific insights, as benchmarked by FIRE-Bench \cite{FireBench}.

Additional Related Works can be found in the Supplement.


\section{Methodology}
\label{sec:method}

\subsection{Data Sources and Pre-processing}

For quality control and preprocessing, all datasets listed in Supplement Table \ref{tab:datasetMetadata} were either obtained directly as processed single-cell datasets (Seurat or AnnData `h5ad` format) or downloaded as raw count matrices and subsequently processed following the original paper's detailed instructions. We normalized total counts per cell using \texttt{scanpy.pp.normalize\_total}, and applied log-transformation with \texttt{scanpy.pp.log1p}. The unsupervised clusters and ground truth annotations were directly borrowed from the metadata information of original datasets.

\subsection{Iterative Annotation Pipeline}

Let $\mathcal{D}$ represent our input single-cell dataset with $N$ cells, where each cell's gene expression is denoted as $\mathbf{x}_i \in \mathbb{R}^M$ for $M$ genes. At each iteration $t$, clusters are represented as $\mathcal{C} = \{C_1, \dots, C_K\}$, with annotations $\alpha_t: \mathcal{C} \rightarrow \mathcal{T} \cup \{\emptyset\}$ mapping clusters to cell types or remaining undefined. Our pipeline iteratively refines cell type annotations through four specialized stages. The detailed prompts are described in Supplement \ref{box:hypothesis}.

\subsubsection{(i) Hypothesis Generation.} This stage formulates and refines hypotheses regarding cell type distributions by analyzing differential gene expression patterns and prior annotation outcomes. Differential expression analysis is performed using the Wilcoxon rank-sum test implemented in Scanpy. For each cluster $k$, we identify and select the top $N$ genes exhibiting the highest average log-fold change ($\text{avg\_log2FC}$). These genes provide the foundation for hypothesis refinement at each iteration. 

\subsubsection{(ii) Marker Selection.} Based on the hypotheses generated, this stage proposes marker genes that can effectively distinguish specific cell types. It dynamically maintains and updates a list of successful and failed marker genes from previous iterations, enabling adaptive refinement of marker gene proposals. The system strategically selects markers to clarify uncertain or previously unresolved clusters. 

\subsubsection{(iii) Expression Analysis.} The expression analysis stage evaluates proposed marker genes by generating dotplots that visualize gene expression distributions across clusters. This analysis allows for verification of marker gene relevance, ensuring that proposed markers effectively discriminate between cell types and highlight distinct expression patterns. 

\subsubsection{(iv) Result Evaluation.} This stage provides a comprehensive evaluation of the expression analysis results at multiple levels: it assesses gene-level specificity, cluster-level marker signatures, and identifies similar cluster pairs along with their distinguishing features. Confidence scores are generated to quantitatively support annotation assignments, pinpointing clusters requiring further analysis. The recommendation for subsequent zoom-in cluster regions is also attached in the evaluation output. The annotation result is dynamically saved in the environment, while annotated h5ad object and the UMAP visualization of annotation are automatically generated in the folder. 

The system also integrates several adaptive heuristics including marker memory, confidence stabilization, cluster relationship analysis, and contamination detection rules (see Supplementary Methods for details).

\subsection{Benchmark Design and Baselines}

To rigorously assess cell type annotation methods, we develop a comprehensive evaluation framework inspired by the ontology-based scoring methodology employed by GPTCelltype. This framework encompasses systematic procedures, including marker gene name standardization, ontology-based cell type mapping, hierarchical annotation structure construction, and quantitative agreement scoring. Such an ontology-driven approach ensures biologically coherent and meaningful comparisons between predicted annotations and established reference standards.

We intentionally excluded large biological language models requiring extensive fine-tuning, such as scBERT and xTrimoGENE, from our benchmark evaluation due to two primary limitations:

\textbf{Dataset Bias}: These models are predominantly pre-trained on datasets from blood and brain tissues (e.g., 71\% of cells in scGPT’s training data). Consequently, they exhibit performance biases when applied to datasets derived from less represented tissues, such as liver, limiting their generalizability.

\textbf{Computational Constraints}: Fine-tuning these models even on relatively small datasets (approximately 10,000 cells) demands substantial GPU resources (e.g., at least 8GB of GPU memory post-optimization). Such computational requirements surpass the capacities commonly available in standard biological laboratories, posing significant practical barriers to adoption.

\subsubsection{Baseline Methods Included}

Our benchmark incorporates four widely used cell type annotation tools, each exemplifying distinct methodologies ranging from database-driven annotation to advanced machine learning approaches:

\textbf{GPTCelltype \cite{gptcelltype}}: This approach leverages a straightforward LLM-based agent that utilizes the GPT-4 model to annotate cell clusters based on their top 10 marker genes. The annotations produced align directly with the original clustering provided.

\textbf{CellTypist \cite{celltypist}}: A machine learning-based classifier employing an extensive reference database of curated cell types. CellTypist offers over 50 pre-trained models tailored to diverse tissue contexts and conditions, providing annotations at the individual cell level rather than relying solely on predefined clusters.

\textbf{CellMarker 2.0 \cite{cellmarker2.0}}: A comprehensive database compiling manually curated, tissue-specific marker genes. It enables annotations by matching the top marker genes from each cluster against this curated reference. Like GPTCelltype, its annotations adhere directly to the original clustering scheme.

\textbf{scTab \cite{Fischer2024scTab}}: A deep learning model is trained on an augmented dataset comprising 22.2 million cells, is designed specifically to improve generalizability through novel data augmentation strategies. scTab provides cell-level annotations independent of initial clustering.

\textbf{Biomni \cite{huang2025biomni}}: Biomni is a LLM–driven framework designed for general-purpose biological reasoning across diverse omics modalities. It integrates large language models with structured biological knowledge and orchestrator-driven tools to perform tasks such as cell type annotation, functional interpretation, and hypothesis generation. 

In contrast, \textbf{CellMaster} integrates a multi-agent LLM structure to comprehensively evaluate marker gene expression patterns within cell clusters. Unlike the baselines, CellMaster’s annotations are based primarily on initial clustering but can be refined iteratively using its integrated "zoom-in" feature for enhanced granularity.

\subsection{Web Application and UI Design}

\begin{figure}[p]
    \centering
    \includegraphics[width=1\linewidth]{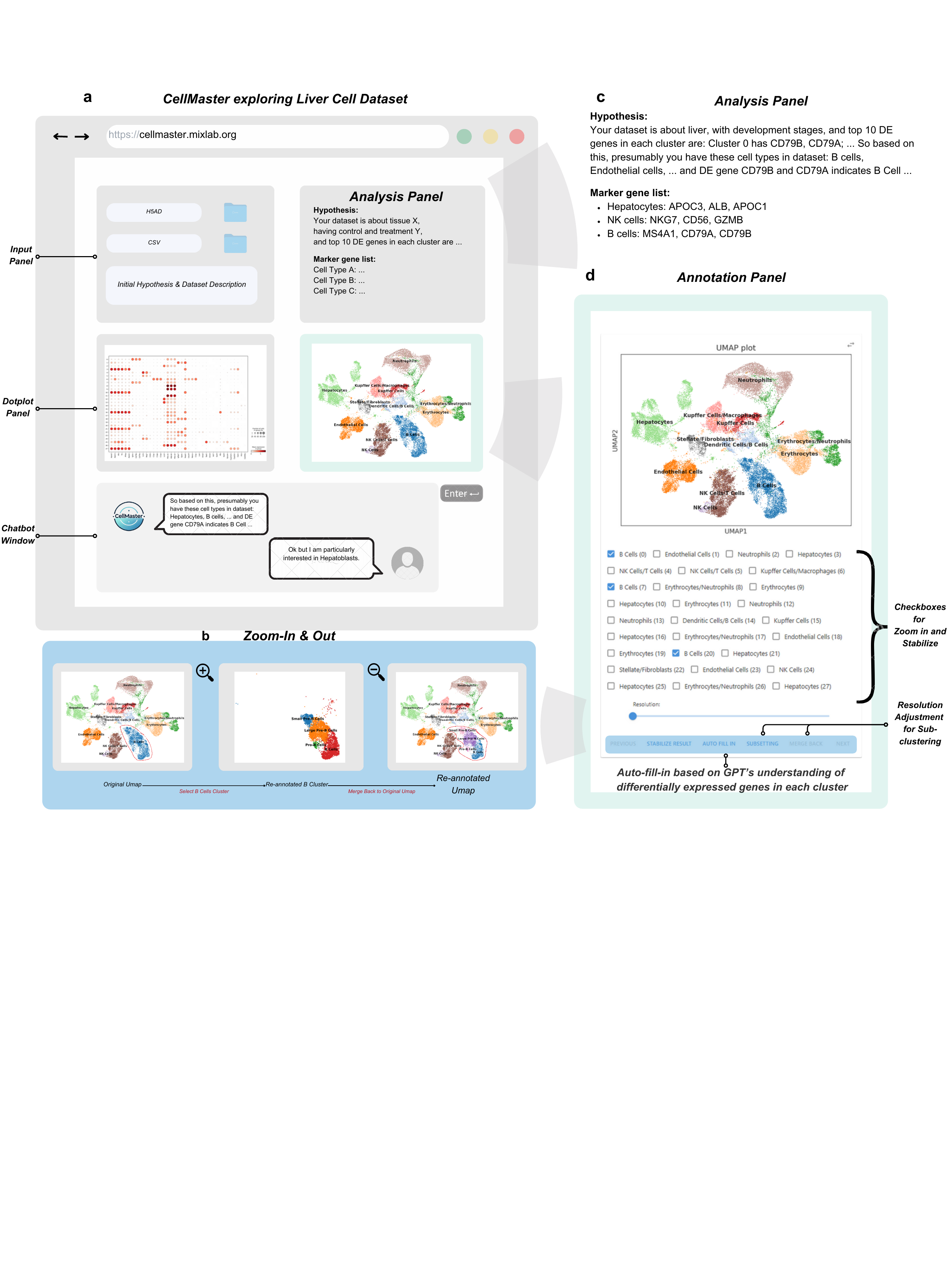}
    \vspace{-225pt} 
    \caption{\textbf{\ours web interface.} 
    \textbf{(a)} Main annotation view with input, dotplot, chatbot, and analysis panels. 
    \textbf{(b)} Zoom-in/out functionality for sub-clustering selected cell populations. 
    \textbf{(c)} Analysis panel showing hypothesis and marker gene lists. 
    \textbf{(d)} Annotation panel with interactive UMAP and cluster controls.}
    \label{fig:ui}
\end{figure}

The web interface for CellMaster is implemented using React, structured around a modular component architecture to enhance usability and facilitate interconnected analytical workflows. The Figure \ref{fig:ui} shows the Panels of the CellMaster UI.

The \textbf{HypothesisInput} component serves as the primary entry point, enabling users to upload single-cell data (H5AD format) and associated metadata through direct file selection or file-path specification. This component incorporates a selector for automated pre-annotations using CellTypist models, alongside an \textit{Explore Cluster} button to initiate analysis upon successful input validation. Additionally, a \textit{Get Other Method} button opens an auxiliary window showcasing alternative annotation methods, including predictions generated by CellTypist and GPT-based annotations.

Visualization of results is centralized within the \textbf{UMAP Plot Panel}, which provides several critical interactive controls:
\begin{itemize}
\item \textbf{Iteration Navigation (Previous/Next):} Navigate through sequential analysis iterations.
\item \textbf{Stabilize Result:} Permanently saves the current cluster annotations to prevent subsequent modifications.
\item \textbf{Auto Fill In:} Automatically annotates unclassified clusters using machine learning-driven predictions.
\item \textbf{Zoom In:} Generates a detailed sub-clustering analysis on selected clusters, offering adjustable resolution (0.1–2.0).
\item \textbf{Zoom Out:} Reverts to the comprehensive dataset view while retaining annotations from sub-cluster analyses.
\item \textbf{Cluster Selection Checkboxes:} Facilitates multi-cluster operations, available only for the latest iteration.
\item \textbf{Swap Icon:} Allows toggling between different UMAP representations within the same iteration when multiple views exist.
\end{itemize}

The \textbf{DotPlot Panel} visualizes gene expression patterns through dot size (representing the proportion of expressing cells) and color intensity (indicating expression levels). Integrated navigation controls allow exploration of expression data across iterations.

The \textbf{AnalysisResults Panel} organizes findings into three specialized tabs:
\begin{itemize}
\item \textbf{Hypothesis Tab:} Presents current biological hypotheses derived from analysis.
\item \textbf{Marker Genes Tab:} Lists proposed marker genes alongside their statistical significance.
\item \textbf{Iteration Summary Tab:} Offers comprehensive conclusions and insights from each analytical iteration.
\end{itemize}

The \textbf{ExploreInput Component} dynamically adjusts to each pipeline stage, providing tailored input fields to facilitate hypothesis refinement, experimental design formulation, contamination detection, and feedback-driven evaluation.

Header controls include:
\begin{itemize}
\item \textbf{Reset Button:} Clears all progress and resets the analysis pipeline.
\item \textbf{Export Button:} Creates and downloads a ZIP archive containing annotated data (H5AD), the complete analysis history, and generated visualizations.
\end{itemize}

Robust state management via React Context ensures seamless synchronization between components. Additionally, workflow orchestration is managed by the \textbf{PipelineManager} class, systematically guiding users through distinct analytical stages, hypothesis formulation, experiment design, environment analysis, rule specification, and evaluation, while providing real-time feedback and comprehensive error handling throughout the annotation process.

\section{Results}

\subsection{\ours architecture and workflow}

\ours is an advanced framework for single-cell RNA sequencing analysis, designed to harmonize automated efficiency with researcher-guided precision. Built upon the widely adopted Scanpy pipeline, the system transcends conventional manual annotation by integrating a multi-agent large language model (LLM) architecture. This innovative approach supports two distinct modes: a fully automated annotation pipeline and a human-in-the-loop interactive workflow. By accommodating both strategies, \ours empowers researchers to align annotations with specific experimental hypotheses, bridging the gap between standardized computational methods and domain-specific biological insights. 

Given a Scanpy object, the system runs an iterative loop: hypothesis generation$\rightarrow$ marker gene proposal $\rightarrow$ evidence synthesis $\rightarrow$ label/abstain; and returns per-cluster labels, natural-language rationales, dotplots, and a confidence score with recommended next steps (e.g., split/merge, add fine-grained markers, run downstream analysis tools). Each step prioritizes biological plausibility and interpretability, synthesizing marker gene expression within the broader experimental context to ensure annotations are both accurate and scientifically meaningful. 

A defining strength of \ours lies in its accessibility and adaptability. Unlike traditional methods requiring computational expertise or reliance on pre-trained reference datasets, this framework democratizes single-cell analysis by eliminating technical barriers. Researchers can freely explore novel datasets without constraints imposed by existing atlases, while dynamically adjusting annotation granularity to suit specific investigative needs, from broad cellular categories to finely resolved subtypes. This flexibility is paired with significant efficiency gains, reducing the time and labor traditionally associated with manual annotation.  

\ours supports two modes. In automatic mode, the system iterates 3 loops as default, yielding a complete label set with confidence scores. In human-in-the-loop mode, experts act as co-pilots via the collaborative UI, reviewing rationales, adding/removing markers, adjusting label granularity, and accepting or revising tentative calls. The UI preserves provenance for all edits and integrates seamlessly with downstream workflows by exporting an annotated Scanpy object. By merging transparency, customization, and computational power, \ours advances single-cell research toward more reproducible, hypothesis-driven discovery.

\subsection{Benchmarking against state-of-the-art tools}

We benchmarked \ours on nine datasets (Liver, PBMC3k, BCL, Myeloid, Large PBMC, Brain, Large Intestine, Limb Muscle, and Retina) against four baseline tools: GPTCelltype, CellTypist, Cell Marker 2.0, and scTab. 
Cluster-level labels were evaluated using a GO-aware similarity rubric (full=1, partial=0.5, mismatch=0), averaged per dataset (macro) following GPTCelltype. Ground truth labels were taken from the original datasets and their corresponding publications.

\subsubsection{Automatic Annotation Benchmarking Performances}

In \ours automatic mode evaluation, we ran \ours's UI for three iterations with only the h5ad dataset and initial hypothesis as input (see Supplementary Information for an example). Unsolved clusters after three iterations were addressed using the ``auto fill in'' function. 

Pretrained models such as scTab and CellTypist showed brittleness under Seurat2/SCT preprocessing, requiring 10k-count normalization in several datasets and failing to match all genes in the Liver dataset. These failure issues illustrated the brittleness of fixed models to out-of-scope data, whereas our LLM-based approach adaptively circumvented such constraints.

As shown in Figure \ref{fig:benchmark}a, across 9 datasets, \ours achieved an average performance of 0.602 $\pm$ 0.058, outperforming best baseline performance of each dataset by 0.071 on average ($\approx13\%$ relative improvement). Notably, Retina (0.705 vs 0.300 - 0.632) and Liver (0.55 vs 0.304 - 0.429) marked the largest gains, while performance remained robust when pretrained baselines failed.

Across five runs per dataset, LLM-based methods showed modest variability (avg. s.d. $\approx$ 0.058 for \ours), with \ours outperforming GPTCelltype on 8/9 datasets, indicating reliable performance across tissues.
We also compared \ours against Biomni (Base Model Gemini-2.5 Pro), another LLM agent based general biology solver. Across representative datasets, \ours consistently outperformed Biomni both in accuracy and stability. On PBMC3k, Retina, and Liver, \ours achieved scores of 0.725, 0.705, and 0.550, whereas Biomni obtained 0.646, 0.570, and 0.464, respectively. Variability was also lower for \ours (s.d. $\leq$ 0.086) compared to Biomni (s.d. up to 0.135). These results highlight that while both systems employ large language models, \ours’s design, integrating structured multi-agent reasoning and iterative feedback, yields more reliable and biologically valid annotations.

\subsubsection{\ours successfully handles challenging edge cases and rare cell groups}

To further elucidate the strengths of \ours, we performed a fine-grained analysis comparing its performance with GPTCelltype and CellTypist. Our analysis focused on three key aspects: annotation granularity (major cell types versus subtypes), performance across broader cell type categories, and the impact of cell and cluster number distinctions. \ours demonstrated significant improvement in sub cell types and small cell groups.

\textbf{Major cell types and subtypes}.
Biologists often annotate datasets with coarse labels for less emphasized cell types and finer subtypes for those of particular interest. For example, in the Liver dataset, T cells are labeled generically as ``T cell,'' whereas in immune cell datasets such as PBMC3k and Zheng68k, T cells are subdivided into CD4 and CD8 T cells, among others. This has been proved important in the recent large-scale atlases of diverse T cell states \cite{TCellNature}. As shown in Figure 3e, \ours significantly outperformed both GPTCelltype and CellTypist in annotating both major cell types and their subtypes.

\textbf{Broader cell type}.
We also evaluated representative broad cell types common in single-cell RNA sequencing datasets (e.g., B cells, T cells, NK cells, and dendritic cells). As shown in Figure \ref{fig:benchmark}e, \ours matched baseline performance for B cells and NK cells while substantially outperforming them for T cells and neural populations. Overall, \ours consistently exceeded GPTCelltype and CellTypist across most broad cell types.

\textbf{Cell and cluster number distinctions}.
We further analyzed the relationship between dataset composition and annotation performance, specifically examining the number of clusters per dataset, the number of cells in the entire dataset, and the number of cells per cluster. As shown in Figure \ref{fig:benchmark}e, \ours consistently outperformed baseline tools across various conditions: when datasets contained more or fewer than 10 cell types, when clusters had fewer than 100, between 100 and 1,000, or more than 1,000 cells, and when the total dataset size exceeded or fell below 45,000 cells. In contrast, CellTypist, when annotating clusters with \textless 100 cells, had especially low performance compared with annotating larger clusters.

\subsection{Human-AI collaboration significantly enhances annotation accuracy}

We evaluated human-in-the-loop \ours under a 3 interaction budget ($\leq$ 3 human inputs each iteration, 1 sentence max for one-time input), where an expert reviewed rationales, supplied brief textual guidance, and acted on uncertainty escalations. Compared to no-feedback (automatic) \ours, accuracy improved by 0.115 $\pm$ 0.085.

The collaborative UI exposes rationales, intermediate step dotplots, confidence scores, "recommend next step" actions, and the result UMAP annotation plot for each iteration. It also records all human inputs / interventions with provenance so accepted and revised labels are fully reviewable in downstream analyses.

\subsubsection{Integration of Manual Annotation Workflow with \ours UI}  

Manual annotation of single-cell data typically requires experts to assemble marker gene lists, examine dot plots, and assign cell-type identities based on domain knowledge. We implemented this workflow directly into the \ours web-based interface, enabling experts to formalize their reasoning, provide iterative feedback, and refine computational predictions without departing from familiar practices.  

The interface operates locally with private model API keys, ensuring data confidentiality, while guiding users through a structured cycle of hypothesis formulation, marker selection, visualization, and iterative refinement. This design preserves expert control while enhancing reproducibility and interpretability.  

\begin{featurebox*}[t]
\begin{tcolorbox}[
  colback=gray!5,
  colframe=gray!50,
  coltitle=white,
  colbacktitle=blue!75!black,
  title=\ours Annotation Workflow,
  fonttitle=\bfseries,
  sharp corners,
  boxrule=0.5mm
]

\textbf{Example Input (Liver Development Dataset)}  
\begin{itemize}
  \item \textbf{Dataset:} 41{,}000 mouse liver cells, 2{,}000 genes, spanning five developmental stages.  
  \item \textbf{Objective:} Characterize hepatocytes, T cells, B cells, and endothelial cells to study liver development.  
\end{itemize}

\textbf{Structured Workflow in \ours}  
\begin{enumerate}
  \item \textbf{Initial Hypothesis:} Provide project background and expected cell types.  
  \item \textbf{Refined Hypothesis:} Elaborate on annotation goals (e.g., highlight specific lineages).  
  \item \textbf{Marker Gene List:} Add or adjust marker genes to tailor reference panels.  
  \item \textbf{Dot Plot Review:} Inspect expression patterns across clusters and flag uncertainties.  
  \item \textbf{Iteration Feedback:} Evaluate and refine annotations until a satisfactory solution is reached.  
\end{enumerate}

\end{tcolorbox}
\caption{Overview of the \ours annotation workflow.}
\label{box:workflow}
\end{featurebox*}

To illustrate the process, an example use case was shown in Box~\ref{box:workflow}, where researchers studying liver development iteratively combined marker-based reasoning with computational outputs. By embedding these steps into an interactive interface, \ours lowered the barrier for adoption and promoted deeper engagement with annotation customization, bridging expert biological insight with scalable computational analysis.

\subsubsection{Human-in-the-loop \ours Consistently Outperforms No-feedback \ours}
\begin{figure}
    \centering
    \includegraphics[width=1\linewidth,trim=0 {0.5\textheight} 0 0,clip, alt={Benchmarking results showing performance heatmap across 9 datasets, head-to-head comparison with Biomni, iteration improvement curves for automate and human-in-the-loop mode, per-dataset comparison, and stratified analysis by cell type and cluster size}]{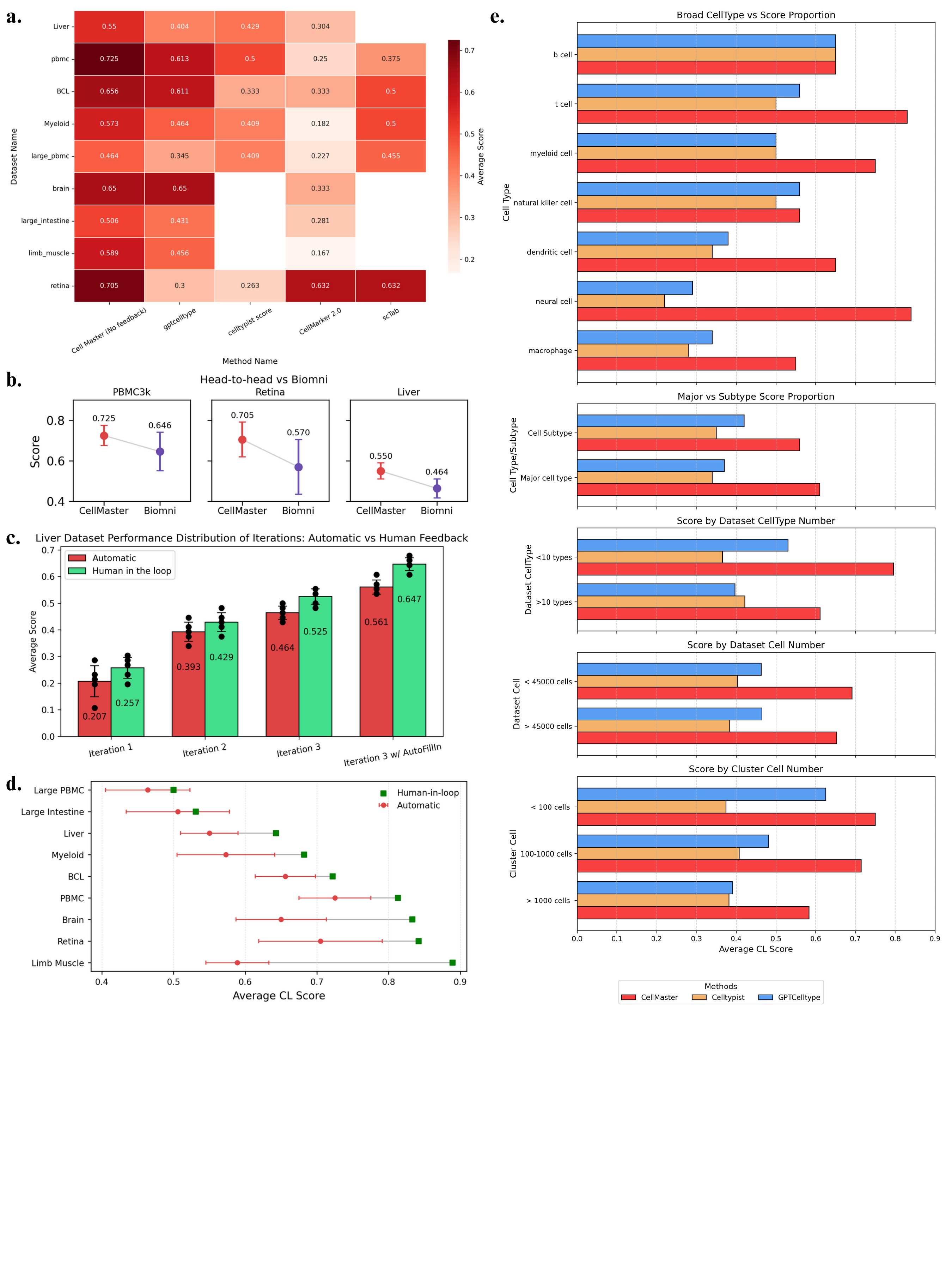}
    \caption{\textbf{Benchmarking results.} 
    \textbf{(a)} Performance heatmap (CL score) comparing \ours against four baselines across 9 datasets; white indicates method failure. 
    \textbf{(b)} Head-to-head comparison with Biomni on three representative datasets. 
    \textbf{(c)} Iterative improvement: automatic vs.\ human-in-the-loop mode on Liver dataset across iterations. 
    \textbf{(d)} Dot plot showing per-dataset performance for automatic (red) and human-in-the-loop (green) modes. 
    \textbf{(e)} Stratified bar chart analysis by cell type category, annotation granularity, dataset size, and cluster size.}
    \label{fig:benchmark}
    \vspace{-13pt}
\end{figure}

In Figure \ref{fig:benchmark}b, we demonstrate that \ours's performance steadily improves with human feedback, achieving an average score margin of 0.115 over the automatic \ours, thus surpassing all the benchmark baselines. In human-in-the-loop mode, the tester utilized the \ours UI to review intermediate step outputs and provided concise feedback, typically in a single sentence. For benchmarking, we opted for general suggestions as input rather than specific requirements for a single cluster or cell type. Additionally, we did not use the zoom-in function in the \ours UI to maintain consistent granularity across the benchmark.
 
Typical human input was brief (e.g., ``Explore additional immune subtypes"), yet sufficient to redirect analysis toward ambiguous regions flagged by the system. \ours reacted with this reasoning: ``In this iteration, we identified immune cells such as T cells. Let's explore the presence of other immune cell types." Such feedback avoided inflation of false positives, showing that minimal human input could substantially boost annotation quality.

Furthermore, in Figure \ref{fig:benchmark}c, we compared the standard deviations of \ours's performance across five experimental runs on the liver dataset in both automatic and human-in-the-loop modes. With three annotation iterations and the auto-fill-in functionality enabled, the human-in-the-loop feedback consistently outperformed the automated approach, demonstrating greater stability and reliability in cell type annotation.

\subsection{Interpreting \ours's decision-making process}

Inspection of rationales across 9 datasets revealed recurring patterns: marker-conflict resolution (e.g., Cd3d vs Nkg7), developmental context usage (e.g., Afp for hepatoblast vs Alb for hepatocyte), and regulon-supported decisions (IRF8 for pDC). In the following case study, we demonstrated how the model continued to improve through iterations without human feedback and traced back the rationale for this enhancement.

\subsubsection{Case Study: Leveraging \ours for Biological Insights}

To demonstrate CellMaster's ability to generate actionable biological insights, we simulated a beginner user analyzing a developmental mouse Liver dataset (41,000 cells) without prior tissue knowledge. Starting from a coarse hypothesis, \ours iteratively proposed labels, flagged uncertain clusters, and suggested zoom-ins on Neutrophils and B cells based on evidence summaries.

The workflow reproduced key findings from the original study and generated additional hypotheses, including novel subsets within neutrophils/B cells supported by marker sets and developmental patterns. These insights were elaborated in the collaborative UI and exported for downstream validation in Figure \ref{fig:casestudy}.

\subsubsection{Annotation Process Simulation} 

\begin{figure}
    \centering
    \includegraphics[width=1\linewidth,trim=0 {0.6\textheight} 0 0,clip, alt={Case study workflow showing Neutrophil subtype resolution in developmental liver, including dotplot of developmental genes, UMAP visualizations, and annotations for immature, intermediate, and mature Neutrophils}]{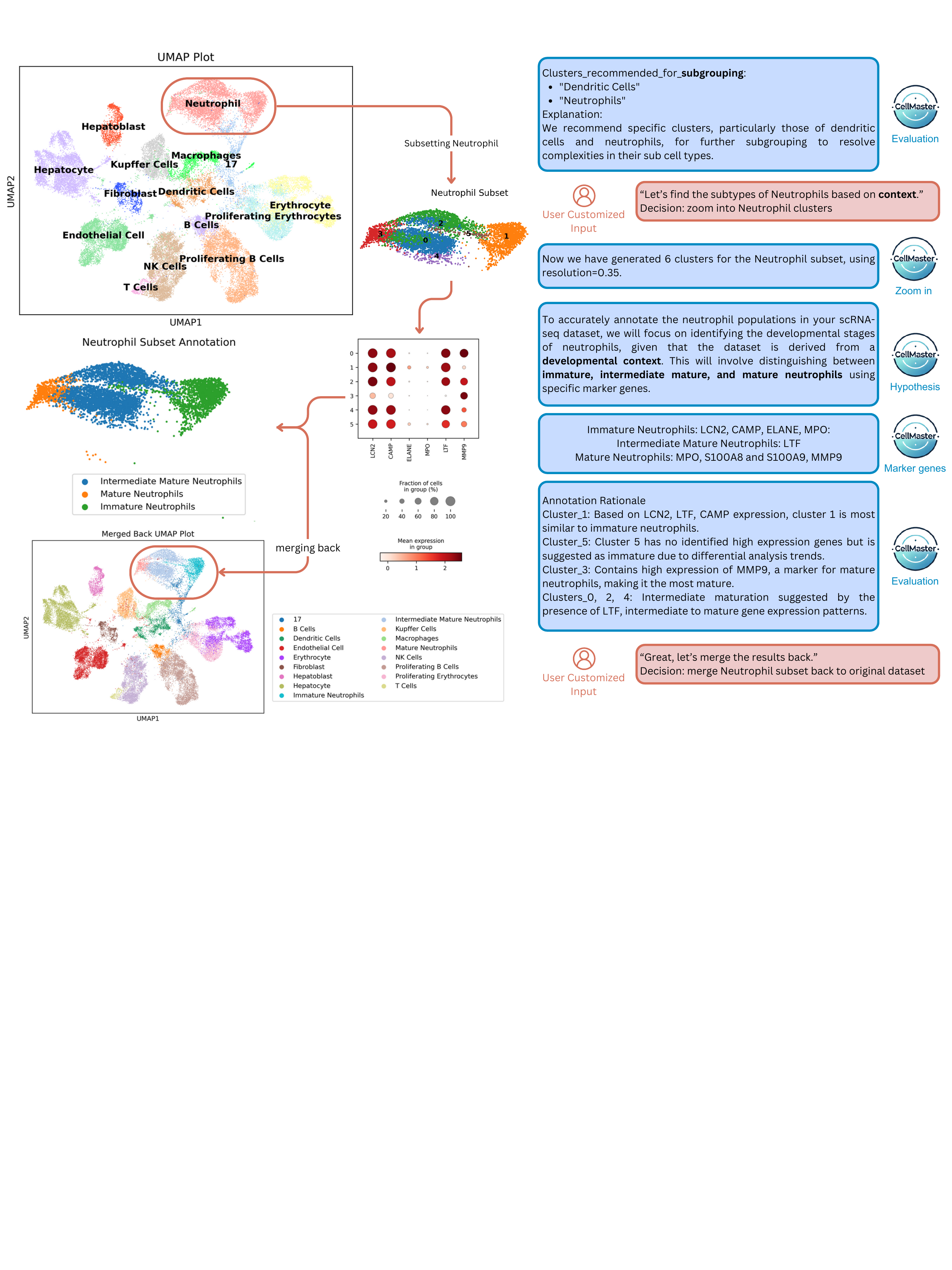}
    \vspace{-120pt}
    \caption{\textbf{Case study: Neutrophil subtype resolution in developmental liver.} 
    \ours recommends sub-clustering Neutrophils, proposes developmental stage markers (LCN2, LTF, MMP9), and provides rationales for immature, intermediate, and mature neutrophil assignments. Final annotations are merged back into the full dataset.}
    \label{fig:casestudy}
    \vspace{-10pt}
\end{figure}

Dataset description and hypothesis were entered as: ``41,000 cells across five developmental stages… likely cell types include T, B, hepatocytes, endothelial…''. Two automatic iterations with one brief instruction, ``Look for complicated clusters and identify their types'', yielded major cell types upleft UMAP in Figure \ref{fig:casestudy}.

In iteration 2, \ours recommended subgrouping Neutrophils, prompting the user to zoom in and generate six sub-clusters.
Leveraging the developmental context, \ours proposed immature, intermediate, and mature neutrophil states supported by marker genes (Lcn2, Ltf, Camp, Mmp9).
Sub-clusters were annotated based on marker strength and developmental stage (Lcn2/Ltf/Camp for immature, partial Ltf for intermediate, and Mmp9 for mature).
The user accepted the rationale, merged the subsets, and obtained an updated UMAP automatically.

\ours's subset annotations aligned with the dataset's developmental context and were supported by rationales, dot plots, and UMAPs. The biologically plausible labels matched the original publication, highlighting \ours’s ability to deliver actionable insights with minimal user input.

\subsubsection{Comparison with baseline and broader implications}

Unlike manual or static marker-based approaches, \ours employs iterative feedback and context-aware reasoning to deliver more nuanced subtype annotations.
Compared to GPTCelltype, which occasionally proposed context-inconsistent neutrophil subtypes, \ours reduced off-target labels by grounding decisions in marker evidence and dataset-specific context.

\paragraph{Additional lineage vignette} 
A parallel analysis of the B‑cell compartment reached similar granularity: after two automated iterations, \ours{} flagged clusters rich in canonical B‑cell markers and guided a zoom‑in that resolved \emph{pro‑B}, \emph{large/small pre‑B}, and \emph{naive B} states
({Supplementary Figure \ref{fig:BCellSubgrouping}}). 
This illustrates cross-lineage generalization of the insight-generation module.

\subsection{Ablation study of \ours}

\subsubsection{\ours base model vs performance}

 To study the difference between large backbone models in \ours, we conducted the ablation study on three datasets: PBMC3k, Liver, and Retina. We selected 4 OpenAI API models: `gpt-4o (original \ours)', `gpt-4o-mini', `o1', `o1-mini' along with 2 Gemini API models: `gemini-2.0-flash', `gemini-2.5-pro'. For each model, we run the standard \ours in automatic mode 3 times on each dataset. These experiments were inherited from scPilot~\cite{gao2025scpilot}, our prior work that established the foundation for \ours's agentic architecture. 

While GPT-4o provided a balanced baseline across all datasets (Supp. Fig. 6g), the o1 model improved accuracy in PBMC3k and Retina but declined in the Liver dataset, suggesting potential over-reasoning on complex data. Lightweight variants (gpt-4o-mini, o1-mini) were less consistent, with occasional competitive results but generally higher variability. The Gemini series showed sufficient performance but did not surpass o1. Overall, gpt-4o remains a reliable default, while o1 demonstrates potential for stronger performance at the cost of robustness.

\subsubsection{Iterational performance shift in automatic \ours}

{Supplement Figure 6} illustrates the iterative improvement of \ours on the Liver dataset, which consists of 28 clusters (labeled 0 to 27) containing a diverse range of hepatocytes and non-parenchymal cells (NPCs). We ran \ours in automatic mode, same as benchmarking.  
The performance improved monotonically from 0.179 (2 correct, 6 partial out of 28) at iteration 1, 0.446 (major cell types identified) at iteration 2, to 0.607 by iteration 4 as the plateau. However, in iteration 5, the score dropped to 0.482 and \ours abstained on the previously annotated clusters.

\textbf{Early iteration improvement}.
\ours exhibits a clear improvement trajectory across multiple iterations, in both automatic and human-in-the-loop modes. By surfacing intermediate step outputs in the user interface, the system enables a transparent, white-box annotation process where users can track refinements as they unfold. This iterative setup not only corrects earlier misclassifications but also sharpens cluster-level accuracy. For instance, in cluster 6, the second iteration corrected an initial mislabeling of ``Kupffer cells/resident macrophages" to the more precise ``Kupffer cells." Similarly, the same cell type may vary in annotation accuracy across clusters: clusters 1, 18, and 23 were all identified as ``Endothelial Cells," yet only cluster 1 achieved full accuracy ($>95\%$), whereas clusters 18 and 23 were only partially correct. Together, these patterns highlight how iterative refinement improves both label precision and interpretability.

\textbf{Decreased \ours performance in additional iterations}.
We observed a performance drop at Iteration 5 in {Supplement Figure 5} and conducted an ablation study on three datasets (PBMC3k, BCL, Liver), each annotated three times. Without human feedback, \ours typically peaked within three to five iterations, after which accuracy stabilized or declined. Early iterations (1–4) progressively resolved unsolved clusters, but once all clusters were labeled, further gains required human input. In automatic mode, however, \ours reconsidered already solved clusters, generated new marker gene lists that sometimes replaced key markers with weaker alternatives. This reduced dot plot interpretability and increases annotation errors.

For instance, in the Liver dataset at iteration 5, the omission of `Afp', a critical hepatoblast marker, caused cluster 16 to be mislabeled as ``Hepatocyte" rather than ``Hepatoblast". Similarly, cluster 14, a mixture of dendritic subtypes, was not definitively resolved. As a result, performance fell to 0.482, below \ours’s peak accuracy.

These findings motivate a conservative early-stop in automatic mode (3 iterations in benchmarking) and reliance on uncertainty escalation to focus expert effort where it adds the most value.

\section{Discussions}
\label{sec:roadmap}

\ours's state-of-the-art performance across diverse datasets demonstrates the transformative potential of LLM-driven adaptive reasoning for cell type annotation. Unlike static models such as CellTypist and CellMarker 2.0, \ours leverages LLM's reasoning to dynamically adapts to novel contexts without retraining, while providing interpretable rationales. Tracing reasoning steps such as Neutrophil subtype differentiation in the Liver case study shows \ours capability to address the ``trust deficit'' common in AI tools. The iterative refinement mechanism, improving scores from 0.196 to 0.643 in the Liver dataset, demonstrates how LLMs can simulate expert-like hypothesis testing, a marked advance over deterministic tools

The 0.115 average performance gain with human feedback highlights the critical role of expert intuition in guiding AI. CellMaster’s UI design, which embeds biologist workflows, enables seamless collaboration, contrasting with `black-box` tools. However, ablation studies reveal performance plateaus and drops in no-feedback mode after 3–5 iterations (Supplementary Figure 6f), underscoring AI's current reliance on human oversight for nuanced decisions and to prevent model overconfidence, a challenge observed in other LLM applications.

\section{Limitations}

While CellMaster is effective, several limitations remain. First, the probabilistic nature of LLMs introduces stochasticity, resulting in slightly higher run-to-run variability (mean s.d. 0.058) than simpler baselines such as GPTCelltype (0.053), despite stabilization mechanisms. Second, dependence on commercial model APIs raises cost and privacy concerns for regulated settings. Third, evaluation based on Cell Ontology mapping may bias assessment against genuinely novel cell states absent from existing hierarchies. Finally, the current framework is limited to unimodal transcriptomics, with multi-omics integration left for future work.

\section{Conclusion}
\label{sec:challenges}

CellMaster redefines cell annotation by synergizing LLMs’ adaptive reasoning with biologist expertise through an intuitive, iterative interface. Its out performance of existing tools (Figure 3a) across major tissues and subtypes, coupled with robust handling of edge cases, positions it as a versatile solution for diverse single-cell studies. The framework’s emphasis on interpretability and human collaboration addresses critical barriers in AI adoption for life sciences. Future work should focus on reducing stochasticity via constrained decoding and expanding ontology coverage to emerging cell types. Future work will focus on reducing stochasticity via constrained decoding, expanding ontology coverage, and integrating multi-modal data including scATAC-seq and spatial transcriptomics. By democratizing expert-level annotation, CellMaster accelerates the transition from data generation to biological insight.

\section{Ethics and Data Use} 

This study relies exclusively on publicly available, anonymized datasets derived from prior publications (see Supplement Table \ref{tab:datasetMetadata}). No new human subjects data were generated, collected, or processed for this work. Consequently, this study constitutes a secondary analysis of de-identified data and is exempt from Institutional Review Board (IRB) oversight.

\section{Data and Code Availability}

All of the processed scRNA-seq datasets used in CellMaster benchmarking are publicly sourced collections. 
The datasets are detailed in Table \ref{tab:datasetMetadata}.
For each dataset, we release the processed objects (H5AD format for scRNA datasets, and csv format for DEG data), along with cell-type labels.  

The complete source code for CellMaster itself, dataset preprocessing, evaluation metrics, and benchmark drivers is released under the MIT license and available at https://github.com/AnonymousGym/CellMaster.

\bibliographystyle{plain}
\bibliography{ref}

\clearpage 

\clearpage

\clearpage

\clearpage

\clearpage

\appendix

\section{Additional Related Works}

\xhdr{Trends in Single-cell Analysis}

The scale of single-cell RNA sequencing (scRNA-seq) has grown exponentially, propelled by large consortia.
such as the Human Cell Atlas (HCA) \cite{humancellatlas}. 
As datasets reach millions of cells, analysis has shifted beyond coarse clustering toward resolving fine-grained and continuous cellular heterogeneity. Cell identity is now understood as a dynamic spectrum shaped by developmental trajectories, microenvironments, and disease states \cite{luecken2019current}. However, capturing such subtle structure remains computationally difficult. Although integration methods such as Harmony \cite{Harmony} and scVI \cite{scVI} effectively mitigate batch effects, they often over-smooth data in latent space, obscuring rare or transient biological states. Preserving biologically meaningful variation while controlling technical noise thus remains a central challenge in large-scale single-cell analysis.

\xhdr{Advancements in Cell Type Annotation}

Cell type annotation underpins most downstream single-cell analyses. Traditional manual annotation, based on expert inspection of marker genes, is interpretable and accurate but inherently unscalable. To address this, reference-based automated methods such as SingleR \cite{singleR} and CellTypist \cite{celltypist} have been widely adopted. While efficient, these approaches are constrained by reference bias: they can only assign labels present in the training atlas, forcing novel or context-specific cell states into existing categories. Moreover, biological cell identities are hierarchical and context-dependent, whereas most automated tools produce flat, fixed labels. This mismatch limits their ability to support hypothesis-driven exploration and nuanced biological interpretation, highlighting the need for more flexible annotation frameworks.

\section{Dataset Information Details}
The benchmark evaluation of CellMaster utilizes nine diverse scRNA-seq datasets spanning eight different tissues to ensure a comprehensive assessment of the system's cross-tissue generalization and performance. These datasets include human and mouse samples such as Liver (mouse liver), PBMC and Large PBMC (human peripheral blood), BCL (human lymphoma), Myeloid (human bone marrow), and Brain, Large Intestine, and Limb Muscle from the Tabula Muris collection. Each dataset was selected to represent varying levels of biological complexity, ranging from the well-defined populations in the 10x PBMC dataset to the intricate developmental stages and fine-grained subtypes found in the mouse liver and human retina datasets. Detailed metadata, including the original source publications and the total cell counts, which range from approximately 2,600 cells to over 65,000 cells, are consolidated in Supplementary Table \ref{tab:datasetMetadata} to provide a clear reference for the benchmark's scope.

\begin{table*}[h]
\centering
\caption{Benchmark dataset metadata and sizes}
\label{tab:datasetMetadata}
\begin{tabularx}{\textwidth}{@{}l l l X@{}}
\toprule
Dataset name      & Tissue              & Source             & Dataset size                  \\ 
\midrule
Liver             & mouse liver         & Developmental Cell Paper \cite{liang2022temporal}    & 41{,}000 cells $\times$ 2{,}000 genes   \\
PBMC              & human PBMC          & 10x Demonstration Dataset \cite{pbmc3k_10x}          & 2{,}638 cells $\times$ 13{,}714 genes    \\
BCL               & human lymphoma      & Cell Discovery Paper \cite{liu2023single}            & 49{,}910 cells $\times$ 25{,}780 genes  \\
Myeloid           & human bone marrow   & CellTypist Demonstration Dataset \cite{xu2023automatic} & 51{,}552 cells $\times$ 36{,}601 genes  \\
Large PBMC        & human PBMC          & Nature Communications Paper \cite{zheng2017massively} & 65{,}943 cells $\times$ 17{,}732 genes  \\
Brain             & mouse brain         & Tabular Muris \cite{tabula2018single}               & 7{,}856 cells $\times$ 23{,}341 genes   \\
Large Intestine   & mouse intestine     & Tabular Muris \cite{tabula2018single}               & 3{,}938 cells $\times$ 23{,}341 genes   \\
Limb Muscle       & mouse muscle        & Tabular Muris \cite{tabula2018single}               & 3{,}900 cells $\times$ 23{,}341 genes   \\
Retina            & human retina        & Nature Communications Paper \cite{menon2019single}  & 20{,}091 cells $\times$ 19{,}719 genes  \\
\bottomrule
\end{tabularx}
\end{table*}

\section{Supplementary Methods}
\subsection{Adaptive Heuristics}

Our system integrates several adaptive mechanisms to enhance annotation accuracy and robustness:

\paragraph{Marker Memory.} To optimize marker gene selection across iterations, the system maintains a record of successful and unsuccessful marker genes. For a gene $g$, its status is determined by:
\begin{equation}
\text{status}_g = \begin{cases}
\text{success}, & \text{if } \exists k: \text{score}_{k,g} \geq 0.7 \cdot \max_{j}(\text{score}_{j,g}) \\
\text{fail}, & \text{otherwise}
\end{cases}
\end{equation}
where $\text{score}_{k,g}$ is a combined measure of expression level and fraction of expressing cells in cluster $k$.
A gene is considered successful if its combined expression score meets or exceeds a numeric threshold, specifically \texttt{success\_threshold = 0.7}. Genes failing to meet this criterion are flagged accordingly, guiding future marker proposals.

\paragraph{Confidence Stabilization.} To ensure stable annotations, the system employs a z-score-based confidence assessment. Annotations are preserved if their corresponding z-scores, computed as normalized expression values, surpass a defined threshold ($Z \geq 2.0$). Clusters that do not meet this confidence criterion are flagged for further refinement in subsequent iterations.

\paragraph{Cluster Relationship Analysis.} 

{\sloppy The pipeline identifies clusters with similar expression profiles to guide targeted refinement through sub-clustering. For clusters $i$ and $j$, similarity is assessed through: \par}
\begin{equation}
\begin{split}
\text{sim}(i,j) = |\{ & g \in G : |\log_2(E_{i,g}) - \log_2(E_{j,g})| > \tau \\
& \text{ or } |F_{i,g} - F_{j,g}| > \delta \}| \leq K
\end{split}
\end{equation}
{\sloppy where $E_{k,g}$ is expression level, $F_{k,g}$ is fraction of expressing cells, $\tau$ and $\delta$ are expression and fraction thresholds respectively, and $K$ is the maximum allowed number of distinguishing genes. 
Similarity between clusters is assessed based on differential expression and fraction thresholds, highlighting pairs of clusters with limited distinguishing genes. This analysis aids in clarifying ambiguous annotations and refining cell type assignments. \par}

\paragraph{Contamination Rules.} The pipeline integrates contamination detection to identify and manage data quality concerns, such as potential doublets and mitochondrial contamination. Specifically, clusters exceeding a predetermined doublet ratio threshold or flagged due to elevated mitochondrial fractions are reviewed, ensuring annotation reliability and dataset integrity.

\section{Evaluation Metrics}

\subsection{Cleaning and Standardizing Cell Type Names}

Raw cell type names often exhibit inconsistencies such as mixed casing, redundant suffixes, or ambiguous abbreviations. To address this, a cleaning and standardization process was applied:

\begin{itemize}
\item \textbf{Automatic Cleaning:} Plural forms (e.g., ``cells'') were singularized, redundant whitespace and punctuation removed, and biologically relevant symbols (e.g., slashes ``/'') retained.
\item \textbf{Standardized Mapping:} Cleaned names were first matched to a predefined dictionary of standard cell type nomenclature. For unmatched names, a language model dynamically generated standardized mappings to handle uncommon or ambiguous terms.
\end{itemize}

This process harmonized all cell type names, enabling consistent downstream analysis.

\subsection{Mapping to Cell Ontology Identifiers}

Standardized names were mapped to terms in the Cell Ontology (CL) to integrate annotations with structured biological knowledge:

\begin{itemize}
\item \textbf{Ontology Querying:} Standardized names were queried against the Cell Ontology via an automated search, retrieving corresponding ontology identifiers (CLIDs) and their higher-level categories.
\item \textbf{Handling Unmapped Names:} Unmapped names were retained for manual review or further processing.
\end{itemize}

This mapping ensured systematic alignment between predicted and reference annotations.

\subsection{Construction of the Ontology Tree}

The hierarchical relationships within the Cell Ontology were leveraged to evaluate lineage-based relationships between predicted and reference annotations:

\begin{itemize}
\item \textbf{Ontology Hierarchy Parsing:} The Cell Ontology was retrieved in OWL format, from which parent-child relationships were extracted.
\item \textbf{Graph Construction:} A directed acyclic graph (DAG) was constructed with nodes representing CLIDs and edges denoting parent-child relationships.
\end{itemize}

This hierarchical structure enabled evaluation beyond direct matches, incorporating extended lineage-based relationships.

\subsection{Ontology-Based Scoring Framework}

Following the ontology-based scoring methodology established by GPTCelltype, a scoring framework incorporating both exact and hierarchical matches was developed:

\begin{itemize}
\item \textbf{Exact Match:} A score of 1.0 was assigned for identical predicted and reference CLIDs.
\item \textbf{Partial Match:} A score of 0.5 was assigned when predicted CLIDs matched parent or child terms of reference CLIDs.
\item \textbf{No Match:} A score of 0.0 was assigned for no ontology overlap.
\end{itemize}

This scoring framework allowed biologically meaningful comparisons while accounting for the hierarchical structure of the Cell Ontology.

\subsection{Biological Context Validation}

To ensure biological relevance, predictions underwent an additional validation step:

\begin{itemize}
\item \textbf{Relative Identification:} Predictions were validated by cross-checking against parent and child terms related to the reference CLIDs within the ontology graph.
\item \textbf{Broad Type Consistency:} Predictions were compared at higher-level ontology categories (e.g., ``immune cells'', ``stromal cells'') to ensure broad biological consistency.
\end{itemize}

\subsection{Summary}

The evaluation methodology systematically integrates name cleaning, dynamic ontology mapping, and hierarchical scoring for biologically accurate assessment. Leveraging both exact and hierarchical ontology matches, the framework enables robust and biologically meaningful evaluation of cell type annotations.

\section{Statistical Analysis}

All statistics were performed in Python 3.11 using Scanpy 1.10, Scipy 1.13, Numpy 1.26 and statsmodels 0.14.  Differential gene‑expression was computed with the two‑sided Wilcoxon rank‑sum test as implemented in \texttt{scanpy.tl.rank\_genes\allowbreak\_groups(method="wilcoxon")}.  Raw P‑values were adjusted for multiple testing with the Benjamini–Hochberg FDR procedure; genes with adjusted p value \textless 0.05 and $| \text{log2 fold change}| > 1$ were considered significant.  Cluster‑level ``confidence scores'' are z‑scores of log‑normalized expression (In Expression Analysis of Iterative Annotation).

Benchmark performance was evaluated with the ontology\-  metric of GPTCelltype.  Agreement scores (exact = 1.0, parent/child = 0.5, no match = 0) were averaged across clusters;  Differences between methods were tested using a paired Wilcoxon signed‑rank test across the nine datasets; FDR was controlled at 5\%.
All tests were two‑sided unless noted.

\section{Comprehensive Feature Comparison with Existing Paradigms}

To rigorously position CellMaster within the current annotation landscape, we evaluated it against three established workflows: (1) Standard Automated Tools (e.g., CellTypist), which offer speed but are limited by fixed references; (2) Generic GPT-based approaches, which provide zero-shot capabilities but lack domain-specific grounding; and (3) Manual Annotation, which remains the gold standard for accuracy but is labor-intensive.

As detailed in Supplementary Table \ref{tab:supp_comparison}, traditional automated tools excel in efficiency but lack the flexibility to handle novel cell states or user-defined granularity without retraining. Conversely, while manual annotation allows for high personalization and depth, it scales poorly with dataset size. CellMaster bridges this gap by combining the ``out-of-the-box" availability of Large Language Models with the user-defined granularity typically reserved for manual curation. Unlike static classifiers, CellMaster allows users to guide the resolution of annotation (e.g., ``Immature Neutrophil" vs. "Neutrophil") through natural language, a feature absent in both standard automated tools and rigid GPT scripts.

\begin{table*}[t]
  \centering
  \resizebox{0.85\textwidth}{!}{ 
  \begin{tabular}{>{\raggedright\arraybackslash}p{7cm}cccc} 
    \toprule
    \textbf{Feature} & \textbf{CellMaster} & \textbf{GPT-based} & \textbf{Manual Annot.} & \textbf{Automated Tools} \\
    \midrule
    \textbf{Ease of Use} \newline \textit{(No specialized biological knowledge required)} & \cmark & \cmark & \xmark & \cmark \\
    \midrule
    \textbf{User-Friendliness} \newline \textit{(No coding/scripting required)} & \cmark & \cmark & \xmark & \xmark \\
    \midrule
    \textbf{Efficiency} \newline \textit{(Low time and effort overhead)} & \cmark & \cmark & \xmark & \cmark \\
    \midrule
    \textbf{Data Flexibility} \newline \textit{(No pre-training dataset limit)} & \cmark & \cmark & \cmark & \xmark \\
    \midrule
    \textbf{Out-of-the-Box Availability} \newline \textit{(No external reference construction)} & \cmark & \cmark & \xmark & \xmark \\
    \midrule
    \textbf{Flexible Granularity} \newline \textit{(Adjustable resolution of annotation)} & \cmark & \xmark & \cmark & \xmark \\
    \midrule
    \textbf{Personalization Options} \newline \textit{(Supports user-defined hypothesis)} & \cmark & \xmark & \cmark & \xmark \\
    \bottomrule
  \end{tabular}
  }
  \caption{\textbf{Detailed feature comparison of annotation paradigms.} We contrast CellMaster with GPT-based prompts (e.g., GPTCelltype), Manual Annotation (standard Seurat/Scanpy workflow), and Automated Tools (e.g., SingleR, CellTypist). \cmark  denotes native support; \xmark denotes lack of support or requirement for significant workaround.}
  \label{tab:supp_comparison} 
\end{table*}

\section{Detailed workflow of CellMaster agents}

Below are the detailed prompts and functions used in the multi-agent framework for CellMaster cell type annotation.

\begin{tcolorbox}[
  enhanced,
  breakable,
  width=\linewidth,
  colback=gray!5,
  colframe=gray!50,
  coltitle=white,
  colbacktitle=blue!75!black,
  title=CellMaster Prompt for Cell‑type Annotation,
  fonttitle=\bfseries,
  sharp corners,
  boxrule=0.5mm,
  before skip=10pt, 
  after skip=10pt,   
  label={box:hypothesis}      
]

\textbf{Hypothesis Generation stage:} 
CellMaster integrates the top marker genes per cluster, dataset context, and (optionally) information from previous iterations to create a hypothesis about the dataset.

\tcbline 
content = f"Top \{len(self.top\_genes)\} differentially expressed genes: \{self.top\_genes\}"
        
if self.reference\_dict:

    content += f"You can refer to the possible cell types of these top genes in this dictionary{self.reference\_dict}"
            
content += f"Current Hypothesis:{self.hypothesis}"
        
if annotation\_dict:

    content += f"The cell type annotation from previous iterations {annotation\_dict}"
            
if no\_gene\_cluster:

    content += f"Clusters without need to be focused on: \{no\_gene\_cluster\}"
            
if iteration\_summary:

    content += f"This is summary of previous iteration annotation, with information of next steps to take. {iteration\_summary}"

system\_role =  "You are a research assistant specializing in cell biology. Based on top differentially expressed genes, previous cell type annotation (if provided), Clusters without need to be focused on (if provided), summary of previous iteration annotation (if provided), and failed genes (if provided), refine the given hypothesis to be more accurate and specific."
\end{tcolorbox}

\begin{tcolorbox}[
  enhanced,
  breakable,
  colback=gray!5,
  colframe=gray!50,
  coltitle=white,
  colbacktitle=blue!75!black,
  title=CellMaster Prompt for Cell type annotation - marker gene proposal,
  fonttitle=\bfseries,
  sharp corners,
  boxrule=0.5mm,
  width=\textwidth,
  label={box:marker_gene},     
]

\textbf{Marker Selection stage:}
CellMaster specifically proposes a marker gene list for the cell types of interest. 

\tcbline

            prompt = f'''
    You are a bioinformatics expert specializing in liver cell annotation. Your task is to propose an experiment for cell type annotation based on the following information:

    Refined hypothesis: {self.hypothesis}

    Instructions:
    
    0. We have already labeled some of clusters, the information is in {annotation\_dict}
    
    1.
    a) The most important unsolved clusters are {no\_gene\_cluster}.
    
    b) Other cell types in {annotation\_dict} is cell types we labeled with high or low confidence. We first need to decide whether we should annotate them again.  
    
    2. For the cell types we have already successfully used, the gene marker list is {successful\_genes}.
    
    For large cell types like hepatocyte or B cell, the remaining clusters might still contain them, but for smaller cell types you don't need to think about them again.
    For the marker genes we already used but did not detect any expression, the list is {failed\_genes}. So you don't need to try these gene and related cell type again.
    
    3. Do not specify the cluster with cell type here. You can just output cell type and related marker genes.
    
    4. Only provide proposal of unlabeled clusters, consider potential overlaps in marker gene expression between cell types:
    
    a) Name of the cell type
    b) 3-5 marker genes

    Output format:
    1. List of cell types with their markers:
    [Cell Type 1]: Gene A; Gene B; Gene C
    [Cell Type 2]
    ...

    2. Python list of all marker genes:
    MARKER\_GENES = ['GeneA', 'GeneB', 'GeneC', $...$]

    Remember: 
    - Be specific and concise in your descriptions.
    - Ensure all cell types have at least 3 marker genes.
    - Include the Python list of all marker genes at the end of your response, don't use any backtick.
    '''

    system\_role =  "You are an AI trained to design scientific experiments based on hypotheses and background information."

\end{tcolorbox}

\begin{tcolorbox}[
  enhanced,
  breakable,
  colback=gray!5,
  colframe=gray!50,
  coltitle=white,
  colbacktitle=blue!75!black,
  title=CellMaster Functions - Expression Analysis,
  fonttitle=\bfseries,
  sharp corners,
  boxrule=0.5mm,
  width=\textwidth,
    label={box:expression_analysis}
]

\textbf{Expression Analysis stage:} CellMaster quantifies and contrasts gene expression programs across identified clusters to derive both broad markers and nuanced distinctions. First, average expression and detection fraction matrices are computed for each gene–cluster pair. High‐confidence markers are nominated by threshold filtering (\texttt{identify\_marker\_genes}), ensuring each candidate meets user‐defined minima for mean expression and prevalence. To harmonize gene‐wise dynamic ranges, CellMaster supports either Z-score normalization (\texttt{zscore\_normalize\_expression}) or min–max scaling (\texttt{min\_max\_scale\_expression}). For cluster–cluster comparisons, log2fold changes are calculated (\texttt{compute\_logfc}), and genes that satisfy combined criteria on expression, detection fraction, and absolute fold change (\texttt{identify\_distinguishing\_markers}) are flagged as distinguishing markers. Finally, by surveying all cluster pairs with a relaxable gene‐difference tolerance (\texttt{find\_similar\_cluster\_pairs}), the module reveals pairs of clusters that share overall transcriptional landscapes but diverge by only a small set of signature genes, facilitating downstream analyses such as targeted sub‐clustering or lineage trajectory inference based on these key transcriptional shifts.

\tcbline 

\noindent\texttt{identify\_marker\_genes} \\ 
Iterates over each cluster (rows of \texttt{dotplot\_data}) and each gene (columns) to select those with mean expression 
$\ge$~\texttt{exp\_thresh} and fraction expressing $\ge$~\texttt{frac\_thresh}. Returns a \texttt{marker\_genes} dictionary mapping clusters to lists of candidate markers.

\vspace{1em}

\noindent\texttt{zscore\_normalize\_expression} \\ 
Applies a per-gene Z-score (zero mean, unit variance) transformation across clusters:  
\[
  \text{normalized}_{g,c} = \frac{\text{expr}_{g,c} - \mu_g}{\sigma_g},
  \quad \text{where } \mu_g,\sigma_g \text{ are the mean and standard deviation of gene } g.
\]

\vspace{1em}

\noindent\texttt{min\_max\_scale\_expression} \\ 
Scales each gene’s expression into [0,1]:  
\[
  \frac{\text{expr}_{g,c} - \min_c(\text{expr}_{g,c})}{\max_c(\text{expr}_{g,c}) - \min_c(\text{expr}_{g,c})}.
\]

\vspace{1em}

\noindent\texttt{compute\_logfc} \\ 
Computes $\log_{2}$ fold-change between two clusters with pseudocount 1:  
\[
  \log_{2}\bigl(\text{expr}_{g,\,\text{cluster1}} + 1\bigr)
  - \log_{2}\bigl(\text{expr}_{g,\,\text{cluster2}} + 1\bigr).
\]

\vspace{1em}

\noindent\texttt{identify\_distinguishing\_markers} \\ 
First computes log-fold change via \texttt{compute\_logfc}, then selects genes satisfying: at least one cluster’s expression 
$\ge$~\texttt{exp\_thresh}, fraction $\ge$~\texttt{frac\_thresh}, and absolute log-fold change $\ge$~\texttt{logfc\_thresh}. Returns a list of genes distinguishing the two clusters.

\vspace{1em}

\noindent\texttt{find\_similar\_cluster\_pairs} \\ 
Examines every unordered pair of clusters, applies \texttt{identify\_distinguishing\_markers} with relaxed thresholds 
(\texttt{exp\_diff\_thresh}, \texttt{frac\_diff\_thresh}, \texttt{logfc\_thresh}), and retains pairs whose number of 
differing genes is $\le$~\texttt{max\_diff\_genes}. Outputs a list of 
$(\text{cluster1},\,\text{cluster2},\,\text{diff\_genes})$ tuples.

\end{tcolorbox}

\begin{tcolorbox}[
  enhanced,
  breakable,
  colback=gray!5,
  colframe=gray!50,
  coltitle=white,
  colbacktitle=blue!75!black,
  title=CellMaster Prompt for Cell type annotation - evaluation,
  fonttitle=\bfseries,
  sharp corners,
  boxrule=0.5mm,
  width=\textwidth,
    label={box:evaluation}
]

\textbf{Result Evaluation stage:} CellMaster analyzes the dotplot expression data, and makes comprehensive evaluation about the expression and thus perform prediction. 

\tcbline 

        content =  f'''
        Context:
        You are working on a single-cell RNA sequencing cell type annotation task. The goal is to identify distinct cell types based on gene expression patterns. You have created a dotplot using a set of marker genes to visualize gene expression across different clusters.

        MUST remember: you should list the cell types here. All cell types you refer to, should be in here. {possible\_cell\_types}

        Data:
        For each cluster, the top genes are in this dictionary: {marker\_genes}
        The clusters cannot find top genes are: {empty\_keys}
        These are the genes that are successfully (highly) expressed in some clusters: {success\_list}
        These are the genes that are failed to express in any clusters: {fail\_list}
        There are some clusters that have similar expression, so we did differential expression analysis for these cluster pairs. The pairs and top differential genes are in: {similar\_clusters\_dict}

        Add-on:
        Duplet and Contamination in dotplot: {duplet\_rule}
        Any other important instructions: {contamination\_rule}

        Please give greatest attention to the Add-on part, if they are provided. Make sure to use them in your analysis.

        Instructions:
        Please analyze the provided data and answer the following questions:

        1. Gene-level analysis:
        a) Which genes are highly expressed in specific clusters? Provide a detailed description.
        b) Are there any genes that show differential expression across clusters (high in some, low in others)?
        c) Are there any genes that are not informative for cell type annotation (low expression across all clusters)?
                You should answer this question based on negation of 1b.

        2. Cluster-level analysis:
        a) Are there any clusters that lack high expression of any marker gene? If so, list the cluster numbers. 
        There are in total {cluster\_size} clusters, index from 0.

        3. Overall assessment:
        a) Based on the gene expression patterns, are there distinct clusters that potentially represent different cell types?
        b) Can you assign specific cell type identities to any of the clusters based on the marker gene expression? If so, provide your cell type annotations. You should only assign one cell type to each cluster. No doublet or contamination here.
        c) To refine the cell type annotation, recommend possible additional cell types.
        d) To refine the cell type annotation, recommend any particular cluster to perform subgrouping.

        4. Confidence assessment:
            a) What are confidence levels of your annotation? Please also assign a confidence score to the process name you selected.
            This score should follow the name in parentheses and range from 0.00 to 1.00. A score of 0.00 indicates the
            lowest confidence, while 1.00 reflects the highest confidence. This score helps gauge how accurately the annotation is. 
            Your choices of confidence score should be a normal distribution (average = 0.5)
            You should consider if the annotation is widely seen, using deterministic wording and following background of dataset.
            For instance, if you label a doublet, or using word "probable", the score should be lower.

            b) Based on confidence levels, what are the annotation results that you want to "stabilize", that is not change in next steps? 
            if {iteration} is 1, you should choose top 1/3 confident clusters. if {iteration} is 2 and beyond, you should choose top 2/3 or more clusters. 
            You can choose this threshold.

        Please provide your answers in a structured JSON format, addressing each question separately using "1a" "2a" etc.

        Remember, this is one iteration cell type annotation for a liver scRNA-seq dataset. Your insights will guide further refinement of the analysis.
        '''
        system\_role = "You are an expert bioinformatician specializing in single-cell RNA sequencing data analysis and cell type annotation."
\end{tcolorbox}

\section{Details of UI Inputs and Outputs: Example}

\begin{figure}[t]
    \centering
    \includegraphics[width=0.65\linewidth, alt={Complete CellMaster web interface screenshot showing pipeline stages, input panel with project description, analysis results with hypothesis and marker genes, gene expression dotplot, interactive UMAP with 28 liver cell clusters, and chat feedback interface.}]{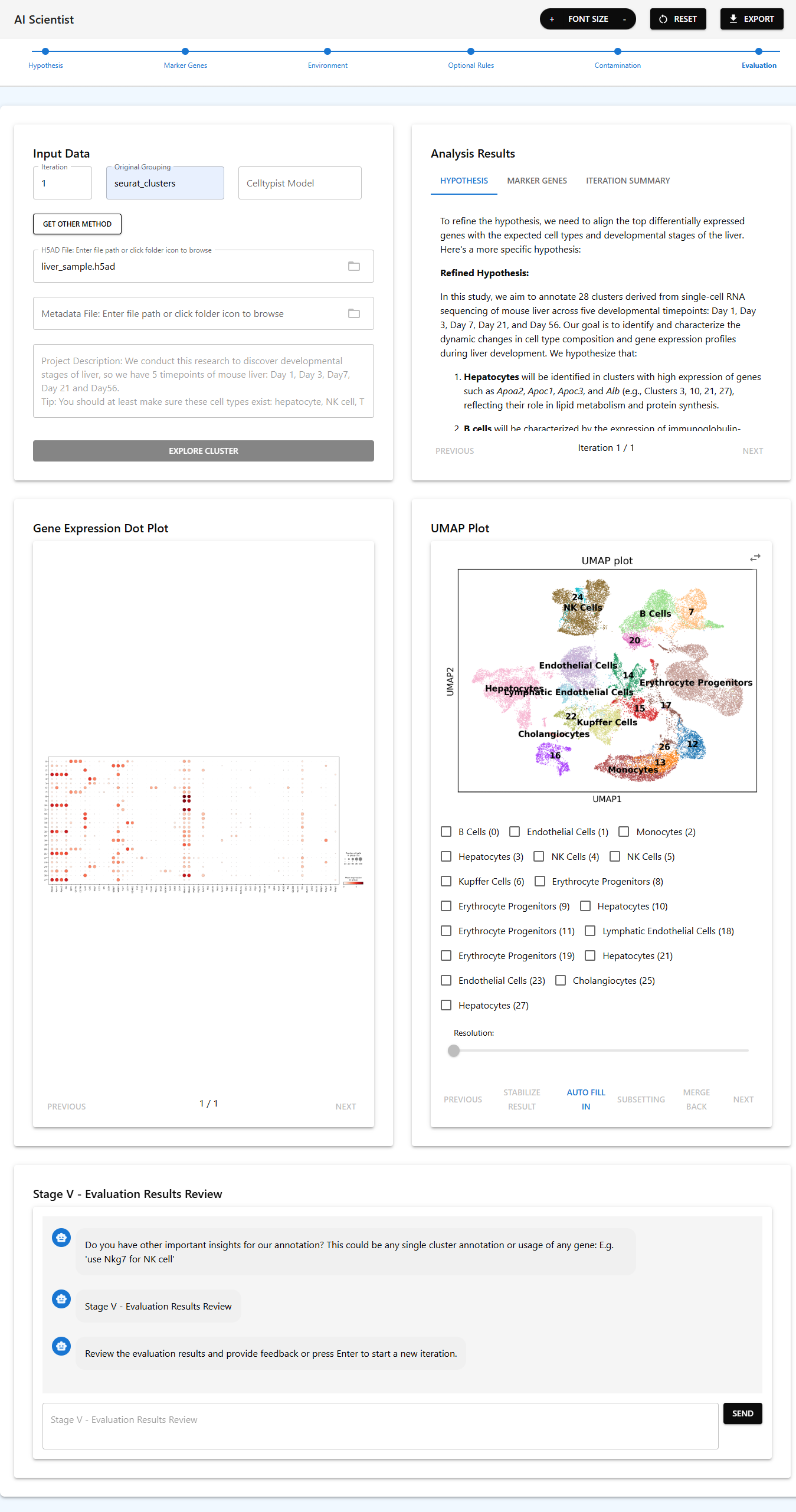}
    \caption{\textbf{Complete \ours web interface example.} Screenshot showing an annotation session on the developmental Liver dataset. The interface includes: (top) pipeline progress indicator and controls; (upper left) input panel for data upload and project description; (upper right) analysis results with hypothesis, marker genes, and iteration summary tabs; (lower left) gene expression dot plot; (lower right) interactive UMAP with cluster selection and sub-clustering controls; (bottom) chat interface for human-in-the-loop feedback.}
    \label{fig:complete_ui}
\end{figure}

Figure~\ref{fig:complete_ui} presents the complete \ours web interface during an annotation session on the Liver dataset. The interface is organized into five main components:

\textbf{Navigation and Controls.} The top bar displays the pipeline progress through six stages (Hypothesis, Marker Genes, Environment, Optional Rules, Contamination, and Evaluation), with the current stage highlighted. Header controls include font size adjustment, a Reset button to restart the analysis, and an Export button to download results.

\textbf{Input Data Panel.} Located in the upper left, this panel accepts user inputs including: iteration number, clustering method selection (e.g., \texttt{seurat\_clusters}), optional CellTypist model for reference annotations, H5AD file path (\texttt{liver\_sample.h5ad}), metadata file, and a free-text project description. In this example, the user describes a developmental liver study spanning five timepoints (Day 1, 3, 7, 21, and 56) with expected cell types including hepatocytes, NK cells, and T cells. The ``Explore Cluster'' button initiates the annotation pipeline.

\textbf{Analysis Results Panel.} The upper right panel displays \ours outputs organized into three tabs: Hypothesis (showing the refined hypothesis with predicted cell types and their marker genes), Marker Genes (listing proposed markers), and Iteration Summary (comprehensive conclusions). The example shows \ours hypothesizing that Hepatocytes will be identified in clusters 3, 10, 21, and 27 based on Apoa2, Apoc1, Apoc3, and Alb expression.

\textbf{Visualization Panels.} The lower section contains two visualization panels. The Gene Expression Dot Plot (left) displays marker gene expression across clusters, with dot size indicating fraction of expressing cells and color intensity indicating expression level. The UMAP Plot (right) shows the annotated dataset with 28 clusters labeled by cell type (e.g., B Cells, Endothelial Cells, Hepatocytes, Kupffer Cells). Checkboxes allow users to select clusters for zoom-in analysis, with a resolution slider (0.1--2.0) controlling sub-clustering granularity. Action buttons include Stabilize Result, Auto Fill In, Subsetting, and Merge Back.

\textbf{Feedback Interface.} The bottom panel provides a chat-style interface for human-in-the-loop interaction. Users can provide feedback such as ``use Nkg7 for NK cell'' to guide subsequent iterations. The system prompts users to review evaluation results and either provide feedback or press Enter to proceed to the next iteration.

\section{Iterative Annotation Performance Dynamics}
\label{supsec:iteration_dynamics}

\begin{figure}
    \centering
    \includegraphics[alt={Ablation study results showing UMAP annotation progression across five iterations on Liver dataset with CL scores from 0.179 to 0.607, performance trajectories for three datasets, and backbone model comparison across six LLM variants.}, width=1\linewidth, trim=0 600pt 0 0, clip]{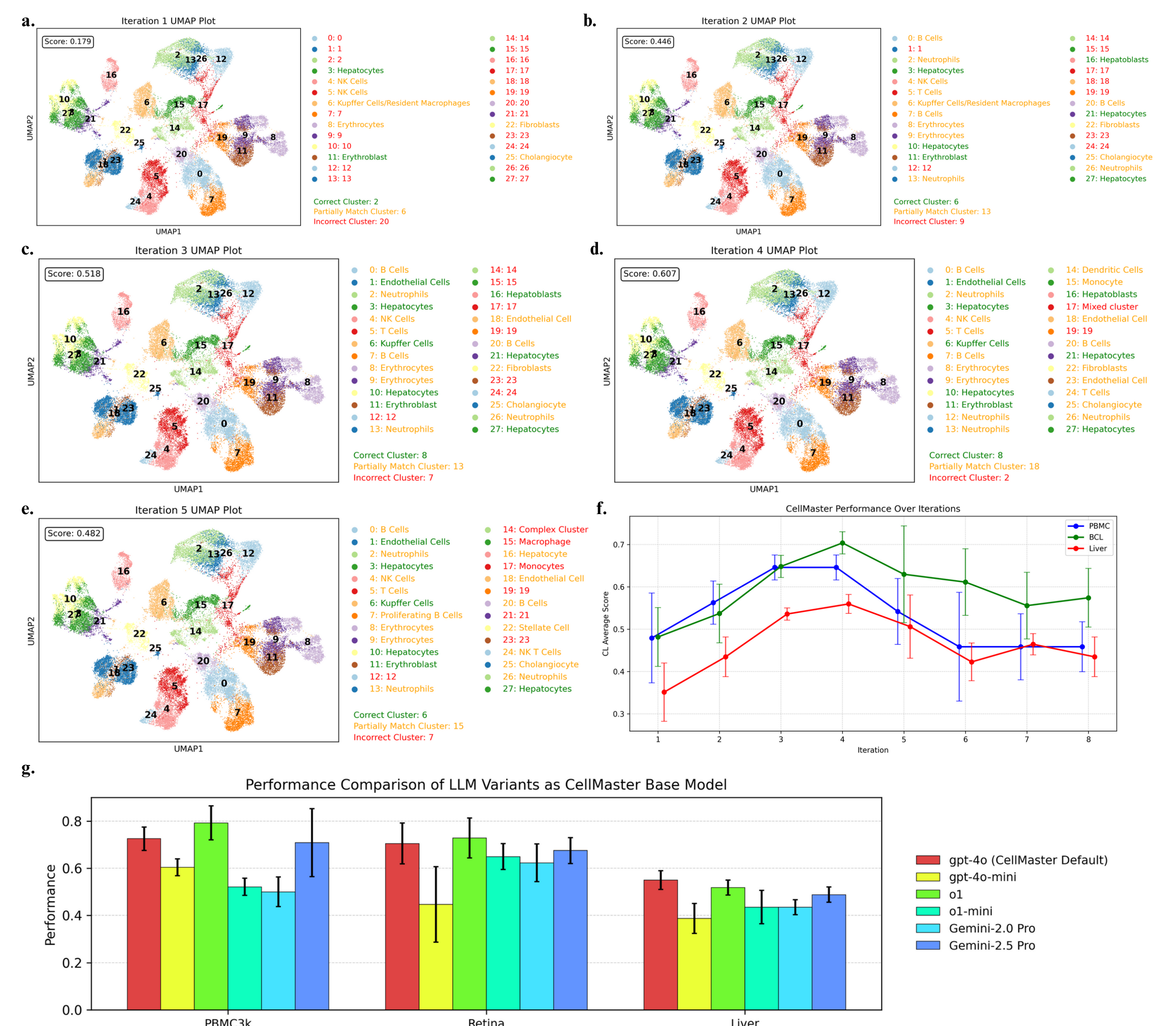}
    \caption{\textbf{Ablation studies: iteration dynamics and backbone model comparison.} 
    \textbf{(a--e)} UMAP visualizations showing \ours annotation progression across five iterations on the Liver dataset, with CL scores improving from 0.179 to 0.607 before declining at iteration 5. 
    \textbf{(f)} Performance trajectories across iterations for PBMC, BCL, and Liver datasets, demonstrating peak accuracy at iterations 3--5. 
    \textbf{(g)} Comparison of six LLM backbone models (gpt-4o, gpt-4o-mini, o1, o1-mini, Gemini-2.0 Pro, Gemini-2.5 Pro) across three datasets. Error bars indicate standard deviation over three runs. Results inherited from scPilot~\cite{gao2025scpilot}.}
    \label{fig:SupplementAblation}
\end{figure}

\textbf{Subfigures~a--e: single–dataset evolution.}  
The first five panels in Figure\ref{fig:SupplementAblation} tracked CellMaster’s fully–automated performance over five successive annotation cycles on the Liver dataset. Accuracy rose monotonically from iteration 1 to iteration 4. This confirmed the agent’s iterative feedback-driven refinements continually annotate blank clusters and correct early mis-labels, incorporating new marker evidence. A noticeable decline occurred at iteration 5, suggesting that, beyond a certain point, the marginal benefit of further automated updates is outweighed by error propagation (e.g.\ compounding low-confidence predictions) and by the diminishing pool of yet-unrefined clusters.

\textbf{Subfigure~f: three–dataset ablation.}  
To verify that this pattern is not dataset–specific, we repeated the analysis on PBMC, BCL, and Liver data (8 iterations each run, 3 runs each dataset). All three curves displayed the same qualitative trend: peak accuracy was reached between the third and fifth iteration, with the exact optimum shifting slightly according to dataset complexity. Prolonged unattended cycling beyond this point produced a mild but consistent drop, mirroring the single–dataset findings.

\textbf{Interpretation and recommended practice.}  
The transient downturn is best understood as a form of \emph{over-iteration}: once high-confidence clusters have stabilized, the agent is left to re-interrogate a shrinking subset of ambiguous cells.  Iterative self-correction without fresh external cues can then amplify spurious signals, eroding global accuracy.  In practice we therefore (i) limit unattended runs to 3 iterations by default in automatic version, (ii) present a ``\textsf{Stabilize Result}'' option in the UI version to lock converged labels, and (iii) encourage a brief manual review before authorizing additional cycles.  These safeguards preserve the demonstrable gains of iterative refinement while preventing the late-cycle regressions observed in Figure~\ref{fig:SupplementAblation}.

\section{Biological insight example of B-Cell subgrouping}
\label{sec:bio_insight_b_cell}

\begin{figure}[h]
    \centering
    \includegraphics[width=1\linewidth, trim=0 750pt 0 0, clip, alt={B-cell subtype resolution workflow showing initial UMAP with B-cell clusters, sub-clustering into 6 groups, dotplot of developmental markers, CellMaster hypothesis and evaluation panels, and final merged UMAP with Pro-B, Large Pre-B, Small Pre-B, and Naive B populations.}]{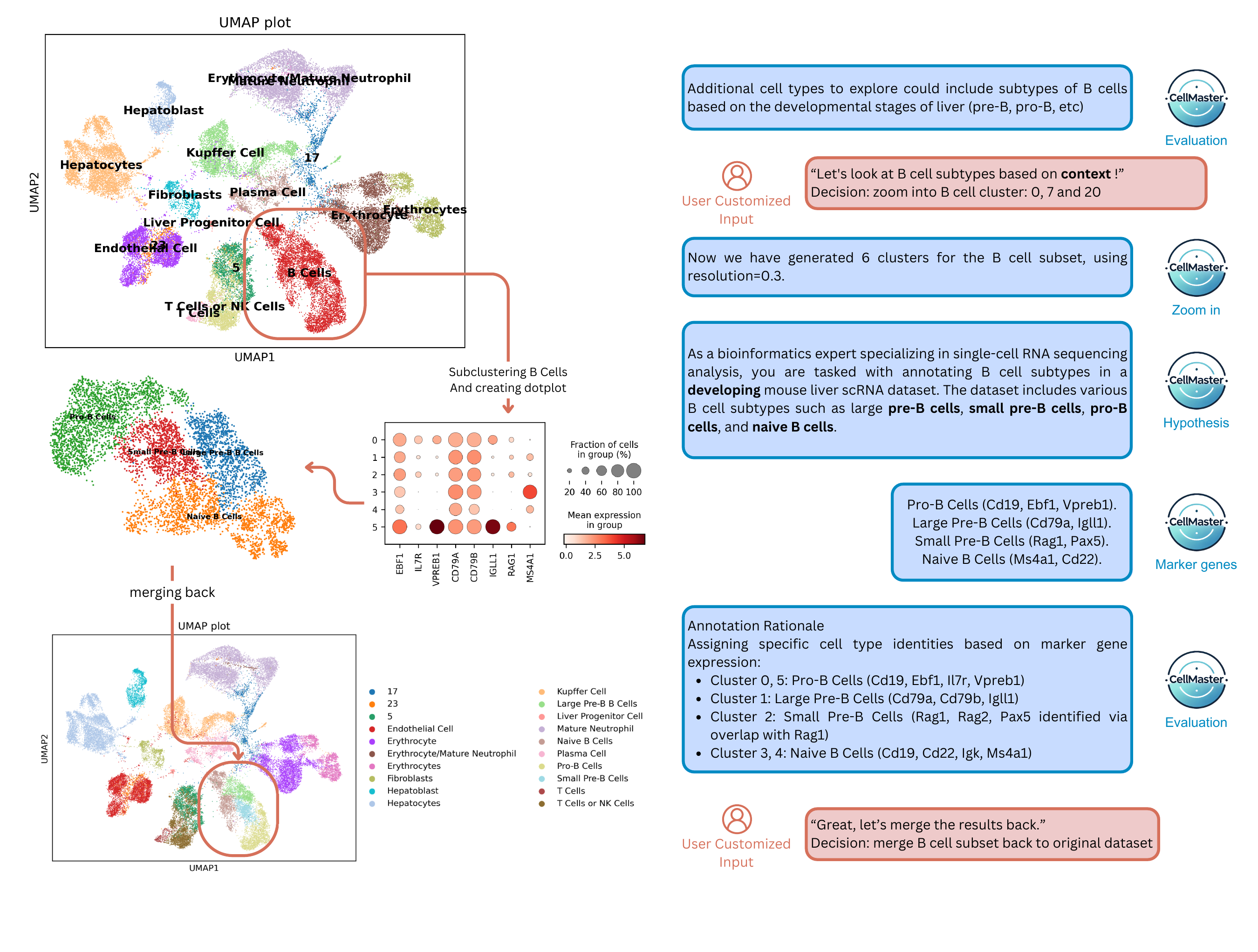}
    \caption{\textbf{B-cell subtype resolution in developmental liver.} Top-left: initial UMAP with B-cell clusters (0, 7, 20) prior to subgrouping. Middle-left: sub-clustered B cells (6 clusters at resolution 0.3). Middle-center: dotplot showing developmental marker expression (Ebf1, Il7r, Vpreb1, Cd79a, Cd79b, Igll1, Rag1, Ms4a1) across sub-clusters. Right panels: \ours workflow including Evaluation recommendation, Hypothesis generation, Marker gene selection, and Annotation rationales. Bottom-left: final merged UMAP with resolved B-cell subtypes (Pro-B, Large Pre-B, Small Pre-B, Naive B) integrated into the full dataset.}
    \label{fig:BCellSubgrouping}
\end{figure}

To demonstrate cross-lineage generalization of \ours's insight-generation module, we performed a parallel analysis on the B-cell compartment in the developmental liver dataset. After initial annotation iterations, the Evaluation module flagged B cells as candidates for further exploration, noting that ``additional cell types to explore could include subtypes of B cells based on the developmental stages of liver (pre-B, pro-B, etc)'' (Figure~\ref{fig:BCellSubgrouping}, top-right, Evaluation panel).

Following user input (``Let's look at B cell subtypes based on context!''), \ours zoomed into B-cell clusters 0, 7, and 20 from the original UMAP (Figure~\ref{fig:BCellSubgrouping}, top-left) and generated 6 sub-clusters at resolution 0.3 (Figure~\ref{fig:BCellSubgrouping}, middle-left, Subclustering panel). The Hypothesis module framed the task around developmental B-cell biology, proposing four canonical stages appropriate for neonatal liver: pro-B, large pre-B, small pre-B, and naive B cells (Figure~\ref{fig:BCellSubgrouping}, right, Hypothesis panel). Corresponding marker genes were selected: Cd19/Ebf1/Vpreb1 for pro-B cells, Cd79a/Igll1 for large pre-B cells, Rag1/Pax5 for small pre-B cells, and Ms4a1/Cd22 for naive B cells (Figure~\ref{fig:BCellSubgrouping}, right, Marker genes panel).

The dotplot visualization (Figure~\ref{fig:BCellSubgrouping}, middle-center) confirmed distinct expression patterns across sub-clusters, with Ebf1 and Vpreb1 enriched in clusters 0 and 5, Cd79a/Cd79b/Igll1 in cluster 1, Rag1 in cluster 2, and Ms4a1 in clusters 3--4. Based on these patterns, the Annotation module assigned subtypes with marker-based rationales (Figure~\ref{fig:BCellSubgrouping}, right, Evaluation panel): clusters 0 and 5 as Pro-B Cells (Cd19, Ebf1, Il7r, Vpreb1); cluster 1 as Large Pre-B Cells (Cd79a, Cd79b, Igll1); cluster 2 as Small Pre-B Cells (Rag1, Rag2, Pax5); and clusters 3 and 4 as Naive B Cells (Cd19, Cd22, Igk, Ms4a1). Upon user confirmation, the resolved subtypes were merged back into the original dataset (Figure~\ref{fig:BCellSubgrouping}, bottom-left), showing distinct Pro-B, Large Pre-B, Small Pre-B, and Naive B populations integrated alongside other liver cell types.

This analysis recapitulates the established B-cell developmental trajectory in fetal and neonatal liver, demonstrating that \ours's reasoning framework generalizes across immune lineages without requiring lineage-specific retraining or marker databases.

\section{Author contributions statement}

Z.W., Y.G., and J.L. conceived the study and designed the core methodology. Z.W. and Y.G. implemented the computational models and performed the primary experiments. J.L. led construction of the UI. J.L. and E.M. processed the datasets and conducted the evaluation analysis. J.C. assisted with software engineering and data visualization. M.A, M.H., J.K., and D.P. provided biological data resources and domain expertise. Z.H. and W.W. contributed to the theoretical framework and provided critical feedback on the machine learning approach. T.I. and E.P.X. supervised the research, and provided overall project guidance. Z.W., Y.G., and J.L. wrote the initial manuscript. All authors reviewed and approved the final manuscript.

\bibliographystylesupp{plain}
\bibliographysupp{supplement_ref}

\end{document}